%% file: main.tex
\documentclass[letterpaper,twocolumn,10pt,pdfusetitle,bookmarks=false]{article}
\PassOptionsToPackage{hyphens}{url}
\usepackage{usenix2021_SOUPS}
\pdfoutput=1
\usepackage{tikz}
\usepackage{pgfplots}
\usepackage{amsmath}
\usepackage[inline]{enumitem}
\usepackage{graphicx}
\usepackage{subcaption}
\usepackage{tcolorbox}
\usepackage{booktabs}
\usepackage{fdsymbol}
\usepackage{csquotes}
\usepackage{tabularx}
\usepackage{multirow}
\usepackage[numbers]{natbib}
\definecolor{orange}{HTML}{E69F00}
\definecolor{skyblue}{HTML}{56B4E9}
\definecolor{bluishgreen}{HTML}{009E73}
\definecolor{yellow}{HTML}{F0E442}
\definecolor{blue}{HTML}{0072B2}
\definecolor{vermillion}{HTML}{D55E00}
\definecolor{reddishpurple}{HTML}{CC79A7}

\definecolor{enthusiastsColor}{HTML}{009E73}
\definecolor{casualsColor}{HTML}{E69F00}

\hypersetup{%
  pdflang={en-US},
  pdftitle={"I have no idea what they're trying to accomplish:" Enthusiastic and Casual Signal Users' Understanding of Signal PINs},
  pdfkeywords={Signal; security; privacy; user study; usability; PINs},
  pdfdisplaydoctitle=true, % For Accessibility
  bookmarksnumbered,
  pdfborder = {0 0 0},
  colorlinks,
  linkcolor={black!80!black},
  citecolor={red!70!black},
  urlcolor={blue!70!black},
  pdfstartview={FitH},
  breaklinks=true,
  hypertexnames=false
}

\usepackage{titlesec}
\begin{document}
%-------------------------------------------------------------------------------

%don't want date printed
\date{}

% make title bold and 14 pt font (Latex default is non-bold, 16 pt)
\title{\Large \bf ``I have no idea what they're trying to accomplish:'' \\ Enthusiastic and Casual Signal Users' Understanding of Signal PINs
\thanks{To appear at Symposium on Usable Privacy and Security 2021}}

% if you leave this blank it will default to a possibly ugly attempt
% to make the contents of the \author command below into a string
\def\plainauthor{Daniel V. Bailey, Philipp Markert, and Adam J. Aviv}

%for single author (just remove % characters)
\author{
{\rm Daniel V. Bailey}\\
Ruhr University Bochum\\
danbailey@sth.rub.de
\and
{\rm Philipp Markert}\\
Ruhr University Bochum\\
philipp.markert@rub.de\\
\and
{\rm Adam J. Aviv}\\
The George Washington University\\
aaviv@gwu.edu
%Name Institution
} % end author

\maketitle

%\thecopyright
%\pagenumbering{gobble}

\input{sections/97-tables}
\input{sections/98-figures}
    
\input{sections/00-abstract}
\input{sections/01-intro}

\input{sections/02-background}

\input{sections/03-related-work}

\input{sections/04-method}
\input{sections/05-results}
\input{sections/06-discussion}

\input{sections/07-conclusion}

\section*{Acknowledgments}
%-------------------------------------------------------------------------------
This material is based upon work supported by the National Science Foundation under Grant No. 184530. Further support was received through the research training group ``Human Centered Systems Security'' sponsored by the state of North Rhine-Westphalia, Germany, and the German Research Foundation (DFG) within the framework of the Excellence Strategy of the Federal Government and the States -- EXC 2092 CASA -- 390781972.

\newpage

%-------------------------------------------------------------------------------
\begin{footnotesize}
\bibliographystyle{plain}
\bibliography{bibliography}
\end{footnotesize} 

\appendix
\input{sections/99-appendix}

%%%%%%%%%%%%%%%%%%%%%%%%%%%%%%%%%%%%%%%%%%%%%%%%%%%%%%%%%%%%%%%%%%%%%%%%%%%%%%%%
\end{document}

%% file: sections/97-tables.tex
\newcommand{\tablePINComposition}[0]{
    \begin{table*}[t]
    \small
    \centering
    \caption{PIN composition across different user groups $n=191$ participants who set a PIN and did not disable it. $t$-tests were performed between groups within categories; all $p$-values are displayed Bonferroni-corrected for 8 overlapping hypothesis tests.}
    \label{tab:PINComposition}
        \resizebox{\textwidth}{!}{
        \begin{tabular}{l | c | cc | cc | cc | cc}
            \toprule
             && \multicolumn{2}{c}{\textbf{Length}} & \multicolumn{2}{c}{\textbf{Digits}} & \multicolumn{2}{c}{\textbf{Letters}} & \multicolumn{2}{c}{\textbf{Special Characters}} \\
            %\cmidrule(lr){3-4} \cmidrule(lr){5-6} \cmidrule(lr){7-8}
            \textbf{Classification} & \textbf{Participants} 
            & \textbf{Mean (SD)} & \textbf{$t$-test}
            & \textbf{Mean (SD)} & \textbf{$t$-test}
            & \textbf{Mean (SD)} & \textbf{$t$-test} 
            & \textbf{Mean (SD)} & \textbf{$t$-test} \\
            \midrule
            Enthusiast & 106 
            & 12.7 \hspace{.5em}(9.8) & $t=4.65$\hspace{1.3em}  
            & 6.2 \hspace{.5em}(3.3) & $t=2.97$\hspace{1.3em} 
            & 4.4 \hspace{.5em}(5.1) & $t=4.57$\hspace{1.3em} 
            & 2.2 \hspace{.5em}(4.1) & $t=2.74$\hspace{1.3em} \\ 
            Casual & \hspace{.5em}85 
            & \hspace{.3em}7.2 \hspace{.5em}(5.7) & $p<0.001^{**}$ 
            & 4.9 \hspace{.5em}(2.4) & $p=0.026^{*}$\hspace{.5em} 
            & 1.4 \hspace{.5em}(3.3) & $p<0.001^{**}$ 
            & 0.9 \hspace{.5em}(2.4) & $p=0.05$\hspace{1.5em} \\ 
            \midrule
            PM User & \hspace{.5em}62 
            & 17.3 (10.2) & $t=9.42$\hspace{1.3em} 
            & 7.0 \hspace{.5em}(3.7) & $t=4.72$\hspace{1.3em}
            & 6.7 \hspace{.5em}(5.1) & $t=8.79$\hspace{1.3em} 
            & 3.7 \hspace{.5em}(4.7) & $t=6.16$\hspace{1.3em} \\ 
            non-PM User & 129 
            & \hspace{.5em}6.8 \hspace{.5em}(5.3) & $p<0.001^{**}$ 
            & 4.9 \hspace{.5em}(2.3) & $p<0.001^{**}$ 
            & 1.3 \hspace{.5em}(3.2) & $p<0.001^{**}$ 
            & 0.6 \hspace{.5em}(2.2) & $p<0.001^{**}$ \\ 
            \midrule
            Overall & 191 
            & 10.3 \hspace{.5em}(8.7) & -- 
            & 5.6 \hspace{.5em}(3.0) & -- 
            & 3.1 \hspace{.5em}(4.7) & -- 
            & 1.6 \hspace{.5em}(3.5) & -- \\
            \bottomrule
        \end{tabular}}
    \end{table*}
}

% enthusiasts vs casuals
% length p < 0.0001, t = 4.6520, df = 189, standard error of difference = 1.203
% digits p = 0.0033, t = 2.9735, df = 189, standard error of difference = 0.429 
% letters p < 0.0001, t = 4.5734, df = 189, standard error of difference = 0.644 
% special p = 0.0067 , t = 2.7441, df = 189, standard error of difference = 0.501 

% pm vs non-pm 
% length p < 0.0001 , t = 9.4280, df = 189, standard error of difference = 1.111 
% digits p < 0.0001, t = 4.7249, df = 189, standard error of difference = 0.458 
% letters p < 0.0001, t = 8.7960, df = 189, standard error of difference = 0.607 
% special p < 0.0001 , t = 6.1612, df = 189, standard error of difference = 0.495 

\newcommand{\tableDemo}[0]{
    \begin{table}[t!]
        \scriptsize
        \centering
        \setlength\tabcolsep{3pt}
        \caption{Demographics of participants divided by subgroups.}\label{tab:demo}
        \resizebox{.95\columnwidth}{!}{
        \begin{tabular}{rrrrrrr}
        \toprule
        & \multicolumn{2}{c}{\textbf{Enthusiasts}} & \multicolumn{2}{c}{\textbf{Casuals}} & \multicolumn{2}{c}{\textbf{Total}} \\
        \cmidrule(lr){2-3} \cmidrule(lr){4-5} \cmidrule(lr){6-7}
        & \textbf{No.} & \multicolumn{1}{c}{\textbf{\%}} 
        & \textbf{No.} & \multicolumn{1}{c}{\textbf{\%}}
        & \textbf{No.} & \multicolumn{1}{c}{\textbf{\%}}\\
        \midrule
        \textbf{Gender} & 132 & 56\,\% & 103 & 44\,\% & 235 & 100\,\%\\
        \midrule
        Male                & 106 & 45\,\% & 71 & 30\,\% & 177 & 75\,\%\\
        Female              &  13 &  6\,\% & 24 & 10\,\% &  37 & 16\,\%\\
        Non-Binary          &   1 &  0\,\% &  1 &  0\,\% &   2 &  1\,\%\\
        Other               &   1 &  0\,\% &  0 &  0\,\% &   1 &  0\,\%\\
        Prefer not to say   &  11 &  5\,\% &  7 &  3\,\% &  18 &  8\,\%\\
        \midrule
        \textbf{Age} & 132 & 56\,\% & 103 & 44\,\% & 235 & 100\,\%\\
        \midrule
        18--24              & 31 & 13\,\% & 14 &  6\,\% &  45 & 19\,\%\\
        25--34              & 57 & 24\,\% & 53 & 23\,\% & 110 & 47\,\%\\
        35--44              & 29 & 12\,\% & 17 &  7\,\% &  46 & 20\,\%\\
        45--54              &  7 &  3\,\% & 10 &  4\,\% &  17 & 7\,\%\\
        55--64              &  5 &  2\,\% &  3 &  1\,\% &   8 & 3\,\%\\
        65--74              &  0 &  0\,\% &  3 &  1\,\% &   3 & 1\,\%\\
        75 or older         &  1 &  0\,\% &  0 &  0\,\% &   1 & 0\,\%\\
        Prefer not to say   &  2 &  1\,\% &  3 &  1\,\% &   5 & 2\,\%\\
        \midrule
        \textbf{Education} & 132 & 56\,\% & 103 & 44\,\% & 235 & 100\,\%\\
        \midrule
        Some High Sch.      &  0 &  0\,\% &  3 &  1\,\% &  3 &  1\,\%\\
        High School         & 31 & 13\,\% & 12 &  5\,\% & 43 & 18\,\%\\
        Some College        &  0 &  0\,\% &  0 &  0\,\% &  0 &  0\,\%\\
        Trade               &  0 &  0\,\% &  4 &  2\,\% &  4 &  2\,\%\\
        Associate's         &  3 &  1\,\% &  6 &  3\,\% &  9 &  4\,\%\\
        Bachelor's          & 35 & 15\,\% & 32 & 14\,\% & 67 & 29\,\%\\
        Master's            & 38 & 16\,\% & 25 & 11\,\% & 63 & 27\,\%\\
        Professional        &  9 &  4\,\% &  3 &  1\,\% & 12 &  5\,\%\\
        Doctorate           & 10 &  4\,\% & 12 &  5\,\% & 22 &  9\,\%\\
        Prefer not to say   &  6 &  3\,\% &  6 &  3\,\% & 12 &  5\,\%\\
        \midrule
        \textbf{Country} & 132 & 56\,\% & 103 & 44\,\% & 235 & 100\,\%\\
        \midrule
        Germany             &  48 &  20\,\% &  20 &   9\,\% &  68 &  29\,\%\\
        USA                 &  25 &  11\,\% &  36 &  15\,\% &  61 &  26\,\%\\  
        United Kingdom      &   7 &   3\,\% &  17 &   7\,\% &  24 &  10\,\%\\
        Other               &  52 &  22\,\% &  30 &  13\,\% &  82 &  35\,\%\\
        \midrule
        \textbf{Background} & 132 & 56\,\% & 103 & 44\,\% & 235 & 100\,\%\\
        \midrule
        Technical           & 96 & 41\,\% & 54 & 23\,\% & 150 & 64\,\%\\
        Non-Technical       & 33 & 14\,\% & 44 & 19\,\% &  77 & 33\,\%\\
        Prefer not to say   &  3 &  1\,\% &  5 &  2\,\% &   8 &  3\,\%\\
        \bottomrule
        \end{tabular}}
%        \vspace{-2em}
    \end{table}
}

%% file: sections/98-figures.tex
\newcommand{\figpromptsall}[0]{  
\vspace{-1em} 
\begin{figure}[h]
%    \begin{figure*}[t]
        \centering
%        \begin{minipage}{0.29\linewidth}
            \centering
            {\includegraphics[width=0.4\columnwidth]{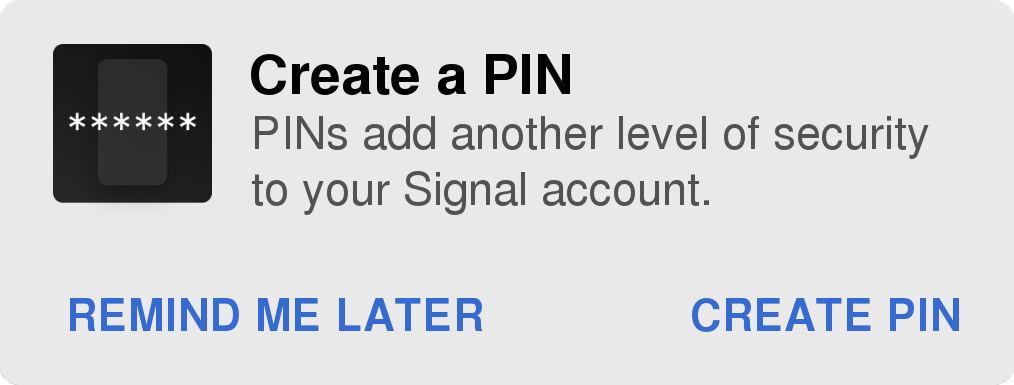}}
            \caption{First prompt to ask Signal users to create a PIN.}
            \label{fig:create:v1}
            %\smallskip
            %{\includegraphics[width=\columnwidth]{figures/verify-pin-big.png}}
            %\caption{Prompt used by Signal to ask users to verify their PIN.}
            %\label{fig:verify}
%        \end{minipage}
\end{figure} 
\vspace{-2em} 
\begin{figure}[h!]        
%        \begin{minipage}{0.29\linewidth}
            \centering
            \tcbox[top=0mm, right=0mm, bottom=0mm, left=0mm, arc=0mm, boxrule=.5pt, colback=white]{
            \includegraphics[width=0.4\columnwidth]{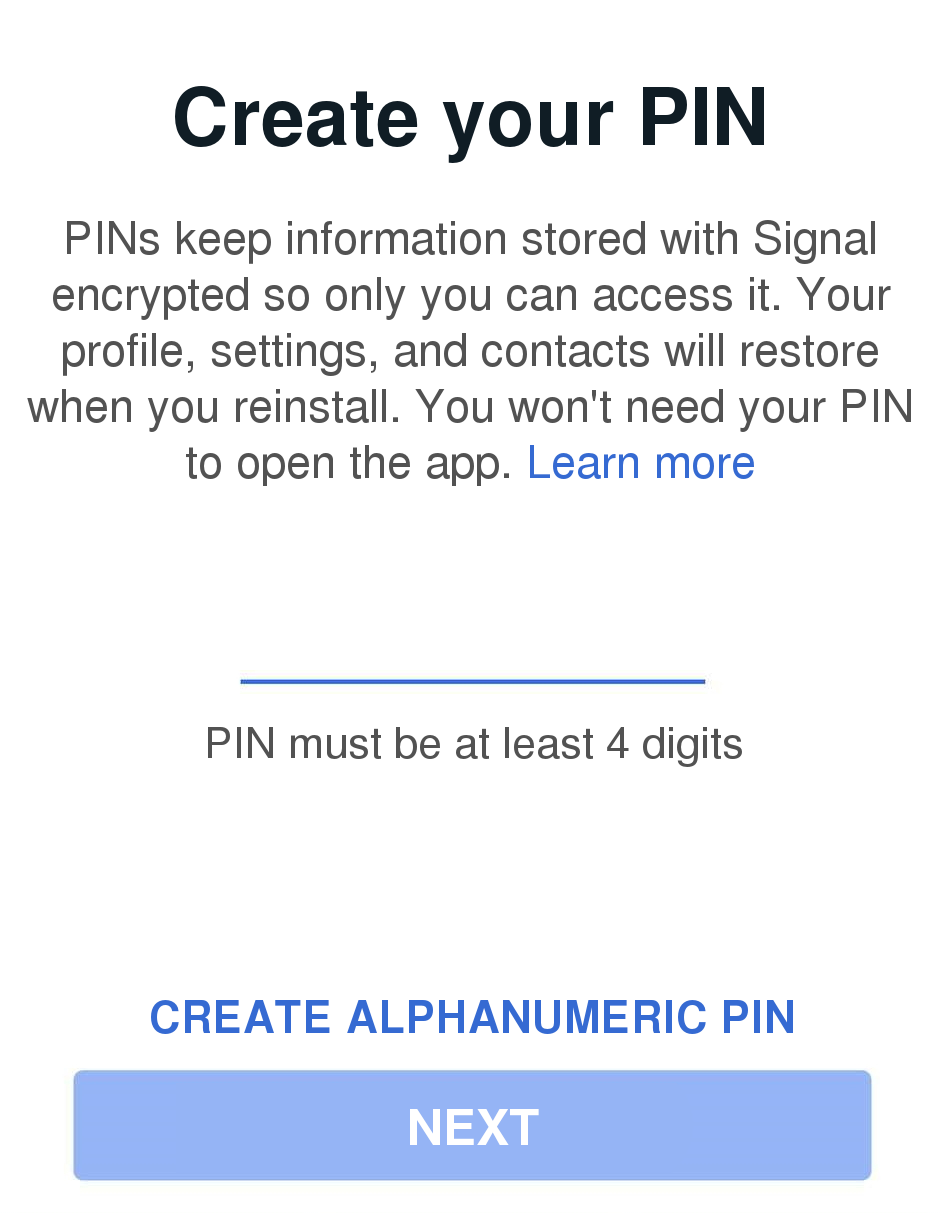}}
            \caption{Updated prompt to ask Signal users to create a PIN.}
            \label{fig:create:v3}
%        \end{minipage}
    \vspace{-1em}
%    \end{figure*}
\end{figure}       
\vspace{-1em} 
%        \hfill
\begin{figure}[h]
%        \begin{minipage}{0.29\linewidth}
            \centering
            \tcbox[top=0mm, right=0mm, bottom=0mm, left=0mm, arc=0mm, boxrule=.5pt, colback=white]{
            \includegraphics[width=0.4\linewidth]{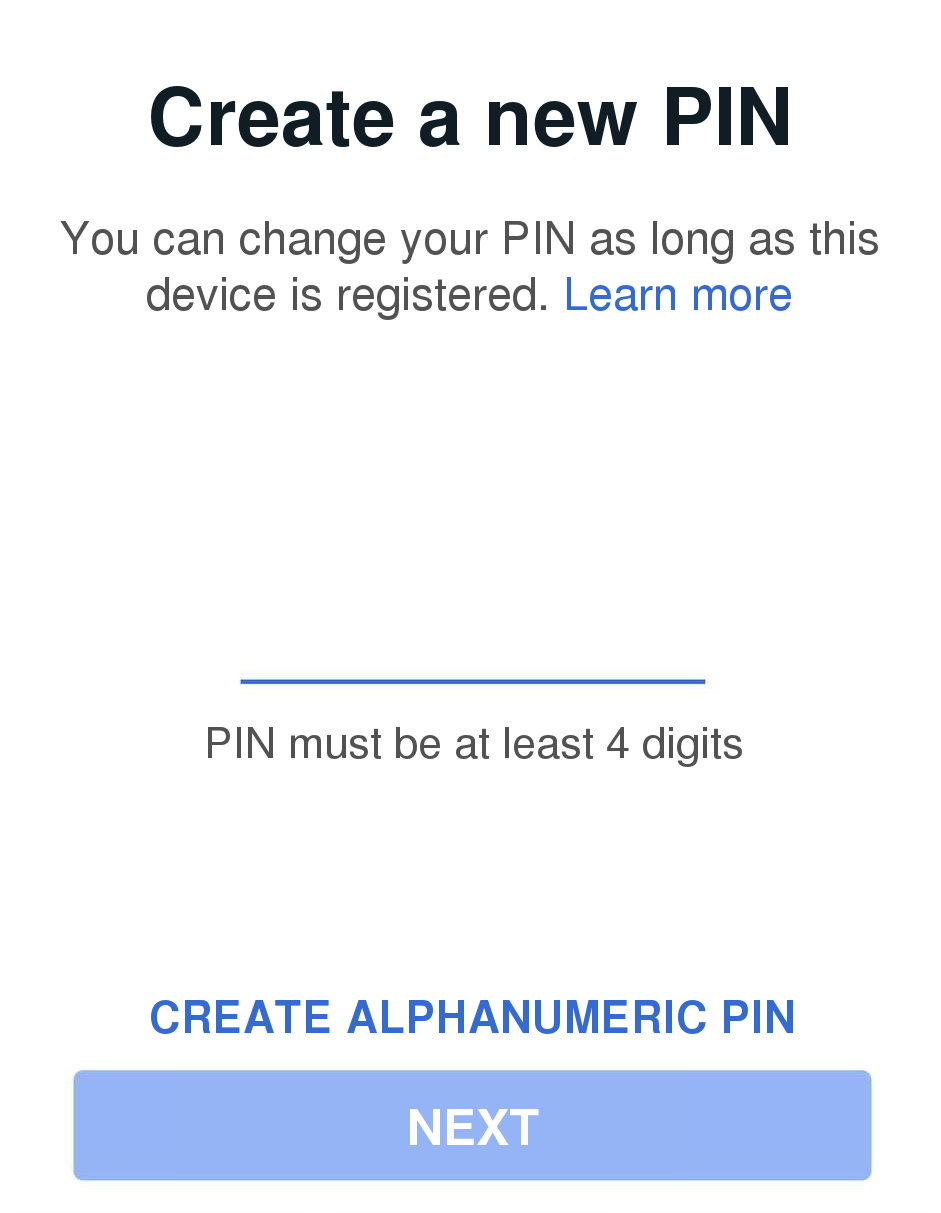}}
            \caption{Prompt used when Signal users wish to change their PIN.}
            \label{fig:create:v2}
%        \end{minipage}
\end{figure}   
}

\newcommand{\figverifypin}[0]{
    \begin{figure}
        \centering
        \includegraphics[width=0.8\columnwidth]{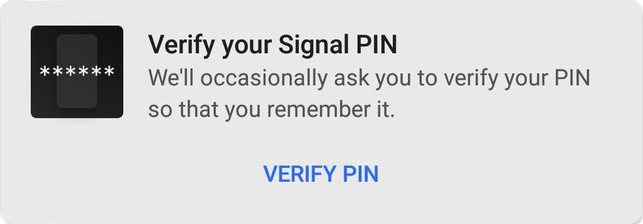}
        \caption{Prompt used by Signal to occasionally ask users to verify their PIN.}
        \label{fig:verify}
    \end{figure}
}

\newcommand{\figfrequency}[0]{
    \begin{figure}
        \centering
        \resizebox{\linewidth}{!}{
        \includegraphics[angle=270]{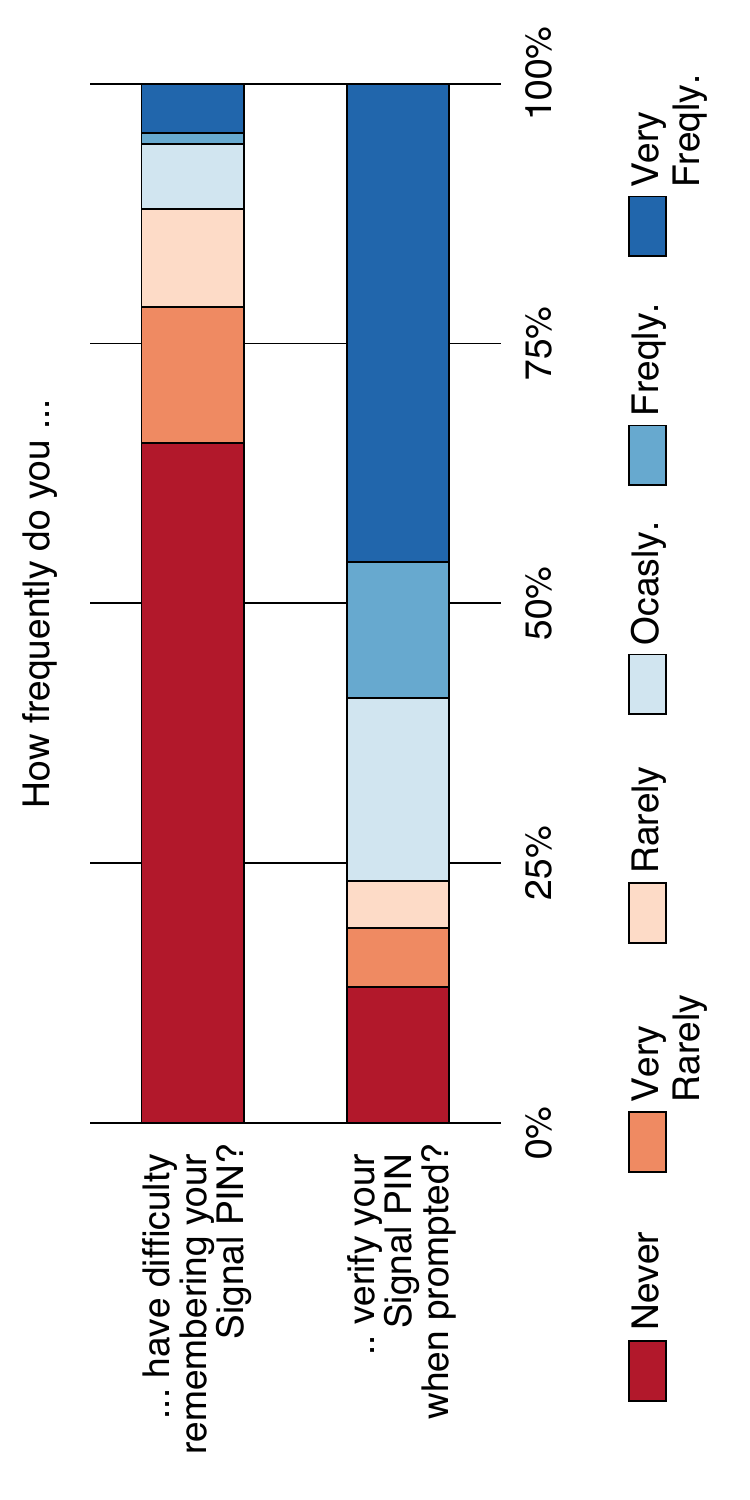}}
        \caption{PIN memorability and verification}
        \label{fig:PINMemorabilityAndVerification}
    \end{figure}
}

\newcommand{\figpromptsthird}[0]{
    \begin{figure}[t]
        \centering
        \begin{minipage}{0.7\linewidth}
        \centering
            \tcbox[top=0mm, right=0mm, bottom=0mm, left=0mm, arc=0mm, boxrule=.5pt, colback=white]{
            \includegraphics[width=0.6\columnwidth]{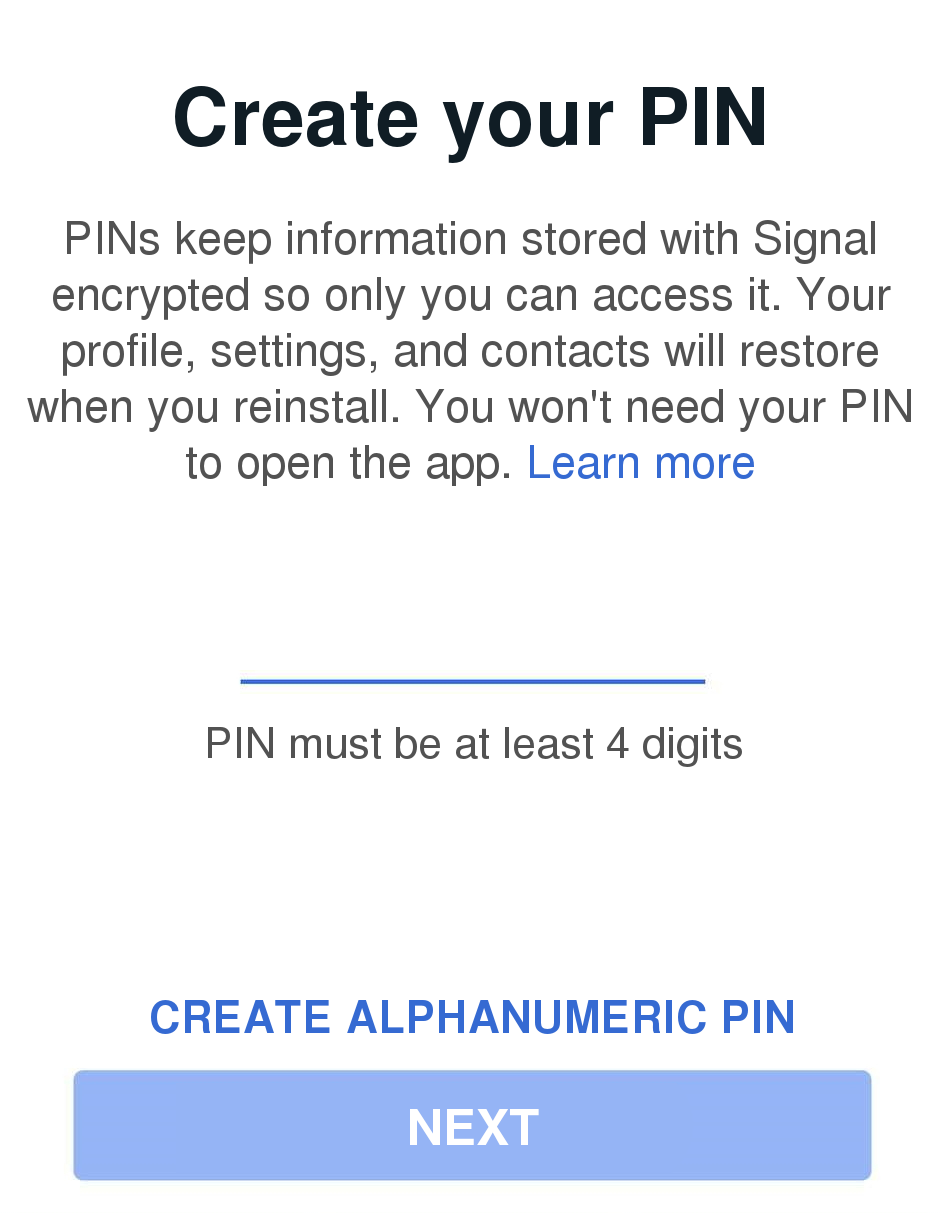}}
            \caption{Prompt used by Signal to ask users to create a PIN.}
            \label{fig:create:v3}
        \end{minipage}
    \vspace{-1em}
    \end{figure}
}

\newcommand{\figUserbreakdownPie}[0]{
\begin{figure}
\centering    
\def\angle{0}
\def\radius{3}
\def\cyclelist{{"orange","blue","red","green"}}
\newcount\cyclecount \cyclecount=-1
\newcount\ind \ind=-1
\resizebox{\linewidth}{!}{
\begin{tikzpicture}[nodes = {font=\sffamily}]
  \foreach \percent/\name in {
      14.0/Reluctant (33),
      48.1/Enthusiasts (113),
      37.9/Casuals (89),
    } {
      \ifx\percent\empty\else               % If \percent is empty, do nothing
        \global\advance\cyclecount by 1     % Advance cyclecount
        \global\advance\ind by 1            % Advance list index
        \ifnum3<\cyclecount                 % If cyclecount is larger than list
          \global\cyclecount=0              %   reset cyclecount and
          \global\ind=0                     %   reset list index
        \fi
        \pgfmathparse{\cyclelist[\the\ind]} % Get color from cycle list
        \edef\color{\pgfmathresult}         %   and store as \color
        % Draw angle and set labels
        \draw[fill={\color!50},draw={\color}] (0,0) -- (\angle:\radius)
          arc (\angle:\angle+\percent*3.6:\radius) -- cycle;
        \node at (\angle+0.5*\percent*3.6:0.7*\radius) {\percent\,\%};
        \node[pin=\angle+0.5*\percent*3:\name]
          at (\angle+0.5*\percent*3.6:\radius) {};
        \pgfmathparse{\angle+\percent*3.6}  % Advance angle
        \xdef\angle{\pgfmathresult}         %   and store in \angle
      \fi
    };
\end{tikzpicture}}
\caption{Participant classification.}
    \label{fig:userbreakdown}
\end{figure}}

\newcommand{\figUserbreakdownBar}[0]{
\begin{figure}
\vspace{1em}
    \centering
    \begin{tikzpicture}
    \begin{axis}[
        xbar stacked,
        area legend,
     	y=8mm, % needed for correct spacing between bars
        legend style={
        legend columns=2,
            at={(xticklabel cs:0.5)},
            anchor=north,
            draw=none
        },
        ytick=data,
        axis y line*=none,
        axis x line*=bottom,
        %tick label style={font=\footnotesize},
        %legend style={font=\footnotesize},
        %label style={font=\footnotesize},
     	every tick/.style={very thin, color=lightgray}, % make sure that ticks look as grid
        xmajorgrids = true, % show grid
        xminorgrids = true, % show minor ticks
     	minor tick num=1, % show minor ticks in between major 
        x axis line style = { opacity = 0 }, % make y axis invisible
        y axis line style = { opacity = 0 }, % make y axis invisible
        ytick style = {draw=none}, % hide y ticks
        yticklabels = {},
        %width=.9\textwidth,
        bar width = 6mm,
        xmin=0,
        xmax=100,
        xticklabel={$\pgfmathprintnumber{\tick}$\,\%},
        enlarge y limits={abs=0.625},
        ]
        \addplot[fill=enthusiastsColor] coordinates
        {
         (56,0)
        };
        \addplot[fill=casualsColor] coordinates
        {
         (44,0) 
        };
        \legend{Enthusiasts, Casuals}
        \coordinate (A) at (28,0);
        \coordinate (B) at (78,0);
    \end{axis}  
    \node at (A) {132};
    \node at (B) {103};
    \end{tikzpicture}
\vspace{.5em}
\caption{Classification of the participants based on the participants' ability to explain the usage of PINs in Signal (\ref{app:survey:q3}).}
\label{fig:userbreakdown-bar}
\vspace{1em}
\end{figure}}

\newcommand{\figUserbreakdownBarNew}[0]{
\begin{figure}[t]
\resizebox{\linewidth}{!}{
    \large
    \begin{tikzpicture}
        \begin{axis}[
        xbar stacked,
        xmin=0,
        xmax=100,
        area legend,
        legend style={
        legend columns=2,
            at={(xticklabel cs:0.5)},
            anchor=north,
            draw=none
        },
        xticklabel={$\pgfmathprintnumber{\tick}$\,\%},
     	y=7mm, % needed for correct spacing between bars
        enlarge y limits={abs=0.45cm}, % needed for correct spacing between bars
     	every tick/.style={very thin, color=lightgray}, % make sure that ticks look as grid
        xmajorgrids = true, % show grid
        xminorgrids = true, % show minor ticks
     	minor tick num=1, % show minor ticks in between major 
        y axis line style = { opacity = 0 }, % make y axis invisible
        ytick style = {draw=none}, % hide y ticks
        symbolic y coords = {
            {Disabled PIN},
            {No PIN Set}},
        ytick=data, % show all names on y axis not only every second
        ]
        
        \addplot[fill=enthusiastsColor] coordinates {
            (45,{Disabled PIN}) %5/11
            (48,{No PIN Set})}; %16/33
        \addplot[fill=casualsColor] coordinates {
            (55,{Disabled PIN}) %6/11
            (52,{No PIN Set})}; %17/33
            \legend{Enthusiasts, Casuals}
            \coordinate (A) at (28,0);
            \coordinate (B) at (78,0);
            \coordinate (C) at (28,100);
            \coordinate (D) at (78,100);
        \end{axis}
        \node at (A) {5};
        \node at (B) {6};
        \node at (C) {16};
        \node at (D) {17};
    
    \end{tikzpicture}}
\vspace{-2em}
\caption{Classification of the participants who disabled or did not set a Signal PIN.}
\label{fig:userbreakdown-barNew}
\vspace{-1em}
\end{figure}}

\newcommand{\figComprehension}[0]{
\begin{figure}[t]
    \begin{subfigure}[htb]{\columnwidth}
    \resizebox{\columnwidth}{!}{
    \begin{tikzpicture}
        \begin{axis}[
        xbar,
        xmin=0,
        xmax=100,
        xticklabel={$\pgfmathprintnumber{\tick}$\,\%},
     	y=6mm, % needed for correct spacing between bars
        enlarge y limits={abs=0.45cm}, % needed for correct spacing between bars
     	every tick/.style={very thin, color=lightgray}, % make sure that ticks look as grid
        xmajorgrids = true, % show grid
        xminorgrids = true, % show minor ticks
     	minor tick num=1, % show minor ticks in between major 
        y axis line style = { opacity = 0 }, % make y axis invisible
        ytick style = {draw=none}, % hide y ticks
        symbolic y coords = {
            {Anti-Cloud},
            {Settings},
            {Registration},
            {Contacts},
            {Encryption},
            {Backup}},
        ytick=data, % show all names on y axis not only every second
        ]
        \addplot[fill=enthusiastsColor] coordinates {
            (8,{Anti-Cloud}) %10/129
            (9,{Settings}) %11/129
            (22,{Registration}) %28/129
            (26,{Contacts}) %34/129
            (33,{Encryption}) %43/129
            (45,{Backup})}; %58/129
        %\coordinate (A) at (4,0);
        %\coordinate (B) at (4.5,6mm);
        %\coordinate (C) at (11,12mm);
        %\coordinate (D) at (13,18mm);
        %\coordinate (E) at (16.5,24mm);
        %\coordinate (F) at (22.5,30mm);
        \end{axis}  
        %\node at (A) {10};
        %\node at (B) {11};
        %\node at (C) {28};
        %\node at (D) {34};
        %\node at (E) {43};
        %\node at (F) {58};
    \end{tikzpicture}}
    \caption{Most popular codes of enthusiasts' responses.}
    \label{fig:enthusiast_responses}
    \vspace{1em}
    \end{subfigure}
    \begin{subfigure}[htb]{\columnwidth}
    \resizebox{\columnwidth}{!}{
    \begin{tikzpicture}
        \begin{axis}[
        xbar,
        xmin=0,
        xmax=100,
        xticklabel={$\pgfmathprintnumber{\tick}$\,\%},
     	y=6mm, % needed for correct spacing between bars
        enlarge y limits={abs=0.45cm}, % needed for correct spacing between bars
     	every tick/.style={very thin, color=lightgray}, % make sure that ticks look as grid
        xmajorgrids = true, % show grid
        xminorgrids = true, % show minor ticks
     	minor tick num=1, % show minor ticks in between major 
        y axis line style = { opacity = 0 }, % make y axis invisible
        ytick style = {draw=none}, % hide y ticks
        symbolic y coords = {
            {\hspace{.75em} Messages},
            {Unlock},
            {Don't Know}},
        ytick=data, % show all names on y axis not only every second
        ]
        \addplot[fill=casualsColor] coordinates {
            (19,{\hspace{.75em} Messages}) %20/106
            (21,{Unlock}) %22/106
            (55,{Don't Know})}; %60/106
        \end{axis}  
    \end{tikzpicture}}
    \caption{Most popular codes of casuals' responses.}
    \label{fig:casuals_responses}
    \end{subfigure}
\caption{Most popular codes of responses to \ref{app:survey:q3}. Please explain how PINs are used by Signal.}
\label{fig:comprehension}
\vspace{-1em}
\end{figure}}

\newcommand{\figWhyPIN}[0]{
\begin{figure}[t]
\resizebox{\linewidth}{!}{
    \begin{tikzpicture}
        \begin{axis}[
        xbar stacked,
        xmin=0,
        xmax=50,
        area legend,
        legend style={
        legend columns=2,
            at={(xticklabel cs:0.5)},
            anchor=north,
            draw=none
        },
        xticklabel={$\pgfmathprintnumber{\tick}$\,\%},
     	y=6mm, % needed for correct spacing between bars
        enlarge y limits={abs=0.45cm}, % needed for correct spacing between bars
     	every tick/.style={very thin, color=lightgray}, % make sure that ticks look as grid
        xmajorgrids = true, % show grid
        xminorgrids = true, % show minor ticks
     	minor tick num=1, % show minor ticks in between major 
        y axis line style = { opacity = 0 }, % make y axis invisible
        ytick style = {draw=none}, % hide y ticks
        symbolic y coords = {
            {Registration},
            {Annoying},
            {Don't know},
            {Required},
            {Prompted},
            {Security}},
        ytick=data, % show all names on y axis not only every second
        ]
        \addplot[fill=enthusiastsColor] coordinates {
            (7,{Registration}) %14/202
            (6,{Annoying}) %12/202
            (2,{Don't know}) %4/202
            (7,{Required}) %14/202
            (4,{Prompted}) %8/202
            (12,{Security})}; %25/202
        \addplot[fill=casualsColor] coordinates {
            (1.5,{Registration}) %2/202
            (3,{Annoying}) %6/202
            (8,{Don't know}) %16/202
            (3,{Required}) %6/202
            (6,{Prompted}) %13/202
            (13,{Security})}; %26/202
        \legend{Enthusiasts, Casuals}
        %\coordinate (A) at (35,0); \coordinate (B) at (78,0);
        %\coordinate (C) at (30,6mm); \coordinate (D) at (75,6mm);
        %\coordinate (E) at (10,12mm); \coordinate (F) at (60,12mm);
        %\coordinate (G) at (35,18mm); \coordinate (H) at (85,18mm);
        %\coordinate (I) at (20,24mm); \coordinate (J) at (70,24mm);
        %\coordinate (K) at (60,30mm); \coordinate (L) at (180,30mm);
        \end{axis}  
        %\node at (A) {14}; \node at (B) {2};
        %\node at (C) {12}; \node at (D) {6};
        %\node at (E) {4}; \node at (F) {16};
        %\node at (G) {14}; \node at (H) {6};
        %\node at (I) {8}; \node at (J) {13};
        %\node at (K) {25}; \node at (L) {26};
    \end{tikzpicture}}
\vspace{-2em}
\caption{Most popular codes assigned to the answers of \ref{app:survey:q5a}. Why did you choose to set a PIN? Only adopters who set a PIN saw this question.}
\label{fig:whypin}
\vspace{-1em}
\end{figure}}

\newcommand{\figWhyNotPIN}[0]{
\begin{figure}[t]
\resizebox{\linewidth}{!}{
    \begin{tikzpicture}
        \begin{axis}[
        xbar stacked,
        xmin=0,
        xmax=50,
        area legend,
        legend style={
        legend columns=2,
            at={(xticklabel cs:0.5)},
            anchor=north,
            draw=none
        },
        xticklabel={$\pgfmathprintnumber{\tick}$\,\%},
     	y=7mm, % needed for correct spacing between bars
        enlarge y limits={abs=0.45cm}, % needed for correct spacing between bars
     	every tick/.style={very thin, color=lightgray}, % make sure that ticks look as grid
        xmajorgrids = true, % show grid
        xminorgrids = true, % show minor ticks
     	minor tick num=1, % show minor ticks in between major 
        y axis line style = { opacity = 0 }, % make y axis invisible
        ytick style = {draw=none}, % hide y ticks
        symbolic y coords = {
            {Memorability},
            {No need},
            {Key management},
            {Anti-Cloud},
            {Inconvenient},
            {Don't know}},
        ytick=data, % show all names on y axis not only every second
        ]
        \addplot[fill=enthusiastsColor] coordinates {
            (3,{Memorability}) %1/33
            (9,{No need}) %3/33
            (9,{Key management}) %3/33
            (15,{Anti-Cloud}) %5/33
            (9,{Inconvenient})}; %3/33
            (6,{Don't know}) %2/33
        \addplot[fill=casualsColor] coordinates {
            (3,{Memorability}) %1/33
            (0,{No need}) %0/33
            (0,{Key management}) %0/33
            (0,{Anti-Cloud}) %0/33
            (12,{Inconvenient})}; %4/33
            (21,{Don't know}) %7/33
        \legend{Enthusiasts, Casuals}
        \end{axis}
    \end{tikzpicture}}
\vspace{-2em}
\caption{Most popular codes assigned to the answers of \ref{app:survey:q5b}: Why did you choose not to set a PIN? Only reluctant who did not set a PIN saw this question.}
\label{fig:whynotpin}
\vspace{-1em}
\end{figure}}

\definecolor{never}{HTML}{b2182b} % purple
\definecolor{very_rarely}{HTML}{ef8a62} % 
\definecolor{rarely}{HTML}{fddbc7} % 
\definecolor{occasionally}{HTML}{d1e5f0} % 
\definecolor{frequently}{HTML}{67a9cf} % 
\definecolor{very_frequently}{HTML}{2166ac} % green

\newcommand{\figLikert}[0]{
\begin{figure}[t]
\centering
\resizebox{\columnwidth}{!}{
\renewcommand*{\arraystretch}{0.7}
    \centering
    \setlength{\tabcolsep}{0pt}
    \begin{tabular}{r c}
& \ref{app:survey:q8}: Difficulty Remembering \\
\raisebox{0.4em}{Enthusiasts\hspace{-.5em}} & 
\begin{tikzpicture}
\begin{axis}[xbar stacked,bar width=1cm,ytick=\empty,y post scale=0.07,xmin=0,xmax=100,ymin=0,ymax=1,legend style={at={(0,-3)},
legend style={cells={align=left, anchor=center, fill}, nodes={inner sep=0.4ex,below=-2ex}},
column sep=0cm, draw=none, anchor=west, legend columns=3},xticklabel style={opacity=0,yshift=12pt}]
\addplot[fill=never, xbar legend, mark=none] coordinates {(67.92,1)}; % 72/106
\addplot[fill=very_rarely, xbar legend, mark=none] coordinates {(16.03,1)}; % 17/106
\addplot[fill=rarely, xbar legend, mark=none] coordinates {(6.60,1)}; %   7/106
\addplot[fill=occasionally, xbar legend, mark=none] coordinates {(3.77,1)}; %   4/106
\addplot[fill=frequently, xbar legend, mark=none] coordinates {(0,1)}; %   0/106
\addplot[fill=very_frequently, xbar legend, mark=none] coordinates {(5.66,1)}; %   6/106
\end{axis}
\end{tikzpicture}\\

\raisebox{0.4em}{Casuals\hspace{-.5em}} & 
\begin{tikzpicture}
\begin{axis}[xbar stacked,bar width=1cm,ytick=\empty,y post scale=0.07,xmin=0,xmax=100,ymin=0,ymax=1,legend style={at={(0.15,-2)},
legend style={cells={align=left, anchor=center, fill}, nodes={inner sep=0.4ex,below=-2ex}},
column sep=0cm, draw=none, anchor=west, legend columns=3},xticklabel style={opacity=0,yshift=12pt}]
\addplot[fill=never, xbar legend, mark=none] coordinates {(63.52,1)}; % 54/85
\addplot[fill=very_rarely, xbar legend, mark=none] coordinates {(10.59,1)}; %  9/85
\addplot[fill=rarely, xbar legend, mark=none] coordinates {(12.94,1)}; %  11/85
\addplot[fill=occasionally, xbar legend, mark=none] coordinates {(9.41,1)}; %   8/85
\addplot[fill=frequently, xbar legend, mark=none] coordinates {(2.35,1)}; %   2/85
\addplot[fill=very_frequently, xbar legend, mark=none] coordinates {(3.53,1)}; %   3/85
\end{axis}
\end{tikzpicture}\\

\\ & \ref{app:survey:q11}: Verify PIN \\

\raisebox{0.4em}{Enthusiasts\hspace{-.5em}} & 
\begin{tikzpicture}
\begin{axis}[xbar stacked,bar width=1cm,ytick=\empty,y post scale=0.07,xmin=0,xmax=100,ymin=0,ymax=1,legend style={at={(0.15,-2)},
legend style={cells={align=left, anchor=center, fill}, nodes={inner sep=0.4ex,below=-2ex}},
column sep=0cm, draw=none, anchor=west, legend columns=3},xticklabel style={opacity=0,yshift=12pt}]
\addplot[fill=never, xbar legend, mark=none] coordinates {(19.42,1)}; % 20/103
\addplot[fill=very_rarely, xbar legend, mark=none] coordinates {(4.85,1)}; %  5/103
\addplot[fill=rarely, xbar legend, mark=none] coordinates {(3.88,1)}; %   4/103
\addplot[fill=occasionally, xbar legend, mark=none] coordinates {(10.68,1)}; %  11/103
\addplot[fill=frequently, xbar legend, mark=none] coordinates {(9.71,1)}; %  10/103
\addplot[fill=very_frequently, xbar legend, mark=none] coordinates {(51.46,1)}; %  53/103
\end{axis}
\end{tikzpicture}\\

\raisebox{5.7em}{Casuals\hspace{-.5em}} & 
\begin{tikzpicture}
\begin{axis}[xbar stacked,bar width=1cm,ytick=\empty,y post scale=0.07,xmin=0,xmax=100,ymin=0,ymax=1,xticklabel={$\pgfmathprintnumber{\tick}$\,\%},
legend style={at={(0,-3.)},
legend style={cells={align=left, anchor=center, fill}, nodes={inner sep=0.4ex,below=-2ex}},
column sep=.1cm, draw=none, anchor=west, legend columns=3, row sep=.1cm}]
\addplot[fill=never, xbar legend, mark=none] coordinates {(4.11,1)}; %  3/73
\addplot[fill=very_rarely, xbar legend, mark=none] coordinates {(6.85,1)}; %  5/73
\addplot[fill=rarely, xbar legend, mark=none] coordinates {(5.48,1)}; %   4/73
\addplot[fill=occasionally, xbar legend, mark=none] coordinates {(27.40,1)}; %  20/73
\addplot[fill=frequently, xbar legend, mark=none] coordinates {(17.81,1)}; %  13/73
\addplot[fill=very_frequently, xbar legend, mark=none] coordinates {(38.36,1)}; %  28/73
\addlegendentry{{\normalsize Never}\hspace{2.6em}};
\addlegendentry{{\normalsize Very Rarely}};
\addlegendentry{{\normalsize Rarely}\hspace{1.7em}};
\addlegendentry{{\normalsize Occasionally}};
\addlegendentry{{\normalsize Frequently}\hspace{.3em}};
\addlegendentry{{\normalsize Very\\ Frequently}};
\end{axis}
\end{tikzpicture}\\

    \end{tabular}}
    \vspace{-1em}
    \caption{PIN memorability and verification.}\label{fig:PINMemorabilityAndVerification}
    \vspace{-1em}
\end{figure}}

\newcommand{\figDisabledReminder}[0]{
\begin{figure}[t]
\resizebox{\linewidth}{!}{
    \begin{tikzpicture}
        \begin{axis}[
        xbar stacked,
        xmin=0,
        xmax=50,
        area legend,
        legend style={
        legend columns=2,
            at={(xticklabel cs:0.5)},
            anchor=north,
            draw=none
        },
        xticklabel={$\pgfmathprintnumber{\tick}$\,\%},
     	y=7mm, % needed for correct spacing between bars
        enlarge y limits={abs=0.45cm}, % needed for correct spacing between bars
     	every tick/.style={very thin, color=lightgray}, % make sure that ticks look as grid
        xmajorgrids = true, % show grid
        xminorgrids = true, % show minor ticks
     	minor tick num=1, % show minor ticks in between major 
        y axis line style = { opacity = 0 }, % make y axis invisible
        ytick style = {draw=none}, % hide y ticks
        symbolic y coords = {
            {No need},
            {Annoyed},
            {PW Manager}},
        ytick=data, % show all names on y axis not only every second
        ]
        \addplot[fill=enthusiastsColor] coordinates {
            (11,{No need}) %5/45
            (11,{Annoyed}) %5/45
            (49,{PW Manager})}; %22/45
        \addplot[fill=casualsColor] coordinates {
            (11,{No need}) %5/45
            (11,{Annoyed}) %5/45
            (2,{PW Manager})}; %1/45
        \legend{Enthusiasts, Casuals}
        \end{axis}
    \end{tikzpicture}}
\vspace{-2em}
\caption{Most popular codes assigned to the answers of \ref{app:survey:q13}: Why did you disable PIN reminders?}
\label{fig:reminder}
\vspace{-1em}
\end{figure}}

\newcommand{\figReuse}[0]{
\begin{figure}[t]
\resizebox{\linewidth}{!}{
    \large
    \begin{tikzpicture}
        \begin{axis}[
        xbar stacked,
        xmin=0,
        xmax=50,
        area legend,
        legend style={
        legend columns=2,
            at={(xticklabel cs:0.5)},
            anchor=north,
            draw=none
        },
        xticklabel={$\pgfmathprintnumber{\tick}$\,\%},
     	y=7mm, % needed for correct spacing between bars
        enlarge y limits={abs=0.45cm}, % needed for correct spacing between bars
     	every tick/.style={very thin, color=lightgray}, % make sure that ticks look as grid
        xmajorgrids = true, % show grid
        xminorgrids = true, % show minor ticks
     	minor tick num=1, % show minor ticks in between major 
        y axis line style = { opacity = 0 }, % make y axis invisible
        ytick style = {draw=none}, % hide y ticks
        symbolic y coords = {
            {\ref{app:survey:q18}: Sharing},
            {\ref{app:survey:q17}: Other apps},
            {\ref{app:survey:q16}: Other contexts},
            {\ref{app:survey:q15}: Screen lock}},
        ytick=data, % show all names on y axis not only every second
        ]
        \addplot[fill=enthusiastsColor] coordinates {
            (1,{\ref{app:survey:q18}: Sharing}) %2/191
            (6,{\ref{app:survey:q17}: Other apps}) %12/191
            (13,{\ref{app:survey:q16}: Other contexts}) %25/191
            (6,{\ref{app:survey:q15}: Screen lock})}; %12/191
        \addplot[fill=casualsColor] coordinates {
            (1,{\ref{app:survey:q18}: Sharing}) %1/191
            (5,{\ref{app:survey:q17}: Other apps}) %9/191
            (15,{\ref{app:survey:q16}: Other contexts}) %28/191
            (7,{\ref{app:survey:q15}: Screen lock})}; %14/191
        \legend{Enthusiasts, Casuals}
        \end{axis}
    \end{tikzpicture}}
\vspace{-2em}
\caption{Frequency of PIN reuse and sharing.}
\label{fig:reuse}
\vspace{-1em}
\end{figure}}

\newcommand{\figStrategy}[0]{
\begin{figure}[t]
\resizebox{\linewidth}{!}{
    \begin{tikzpicture}
        \begin{axis}[
        xbar stacked,
        xmin=0,
        xmax=50,
        area legend,
        legend style={
        legend columns=2,
            at={(xticklabel cs:0.5)},
            anchor=north,
            draw=none
        },
        xticklabel={$\pgfmathprintnumber{\tick}$\,\%},
     	y=7mm, % needed for correct spacing between bars
        enlarge y limits={abs=0.45cm}, % needed for correct spacing between bars
     	every tick/.style={very thin, color=lightgray}, % make sure that ticks look as grid
        xmajorgrids = true, % show grid
        xminorgrids = true, % show minor ticks
     	minor tick num=1, % show minor ticks in between major 
        y axis line style = { opacity = 0 }, % make y axis invisible
        ytick style = {draw=none}, % hide y ticks
        symbolic y coords = {
            %{Word},
            {Pattern},
            {Security},
            {Meaning},
            {Random},
            {Reuse},
            {PW Manager},
            {Memorability},},
        ytick=data, % show all names on y axis not only every second
        ]
        \addplot[fill=enthusiastsColor] coordinates {
            %(1,{Word}) %2/191
            (2,{Pattern}) %3/191
            (2,{Security}) %3/191
            (3,{Meaning}) %6/191
            (8,{Random}) %15/191
            (8,{Reuse}) %16/191
            (15,{PW Manager}) %28/191
            (12,{Memorability})}; %23/191
        \addplot[fill=casualsColor] coordinates {
            %(2,{Word}) %3/191
            (2,{Pattern}) %4/191
            (4,{Security}) %8/191
            (3,{Meaning}) %6/191
            (4,{Random}) %7/191
            (7,{Reuse}) %13/191
            (3,{PW Manager}) %6/191
            (16,{Memorability})}; %30/191
        \legend{Enthusiasts, Casuals}
        \end{axis}
    \end{tikzpicture}}
\vspace{-2em}
\caption{Most popular codes assigned to the answers of \ref{app:survey:q20}: What was your primary strategy in selecting your Signal PIN?}
\label{fig:strategy}
\vspace{-1em}
\end{figure}}

\newcommand{\figSecurityLevel}[0]{
\begin{figure}[t]
\resizebox{\linewidth}{!}{
    \begin{tikzpicture}
        \begin{axis}[
        xbar stacked,
        xmin=0,
        xmax=50,
        area legend,
        legend style={
        legend columns=2,
            at={(xticklabel cs:0.5)},
            anchor=north,
            draw=none
        },
        xticklabel={$\pgfmathprintnumber{\tick}$\,\%},
     	y=7mm, % needed for correct spacing between bars
        enlarge y limits={abs=0.45cm}, % needed for correct spacing between bars
     	every tick/.style={very thin, color=lightgray}, % make sure that ticks look as grid
        xmajorgrids = true, % show grid
        xminorgrids = true, % show minor ticks
     	minor tick num=1, % show minor ticks in between major 
        y axis line style = { opacity = 0 }, % make y axis invisible
        ytick style = {draw=none}, % hide y ticks
        symbolic y coords = {
            {PW Manager},
            {Reuse},
            {Trade-off},
            {Consistent},
            {Memorability},
            {Enough},
            {Security},},
        ytick=data, % show all names on y axis not only every second
        ]
        \addplot[fill=enthusiastsColor] coordinates {
            (3,{PW Manager}) %6/191
            (4,{Reuse}) %7/191
            (5,{Trade-off}) %9/191
            (6,{Consistent}) %11/191
            (6,{Memorability}) %12/191
            (8,{Enough}) %16/191
            (13,{Security})}; %25/191
        \addplot[fill=casualsColor] coordinates {
            (1,{PW Manager}) %2/191
            (1,{Reuse}) %2/191
            (1,{Trade-off}) %2/191
            (3,{Consistent}) %6/191
            (9,{Memorability}) %18/191
            (8,{Enough}) %15/191
            (11,{Security})}; %20/191
        \legend{Enthusiasts, Casuals}
        \end{axis}
    \end{tikzpicture}}
\vspace{-2em}
\caption{Most popular codes assigned to the answers of \ref{app:survey:q22}: Why did you choose a PIN with this security level?}
\label{fig:securitylevel}
\vspace{-1em}
\end{figure}}

%% file: sections/00-abstract.tex
%-------------------------------------------------------------------------------
\begin{abstract}
%-------------------------------------------------------------------------------
We conducted an online study with $n = 235$ Signal users on their understanding and usage of PINs in Signal.  
In our study, we observe a split in PIN management and composition strategies between users who can explain the purpose of the Signal PINs (56\,\%; enthusiasts) and users who cannot (44\,\%; casual users). Encouraging adoption of PINs by Signal appears quite successful: only 14\,\% opted-out of setting a PIN entirely. 
Among those who did set a PIN, most enthusiasts had long, complex alphanumeric PINs generated by and saved in a password manager. Meanwhile more casual Signal users mostly relied on short numeric-only PINs. 
Our results suggest that better communication about the purpose of the Signal PIN could help more casual users understand the features PINs enable (such as that it is not simply a personal identification number). This communication could encourage a stronger security posture.

\end{abstract}

%% file: sections/01-intro.tex
%-------------------------------------------------------------------------------
\section{Introduction}
%-------------------------------------------------------------------------------
\label{sec:introduction}

Signal is an encrypted messaging application that is dedicated to preserving the privacy of its users and enacts features along those lines, such as not centrally storing  users' contact lists, messages, or location histories unencrypted. Signal has historically relied only on users' telephone numbers for identification, authentication (via SMS),  and contact discovery. Unfortunately, these methods are insufficient against attacks, including SIM-swapping~\cite{andrews-18-your-digits, jover-20-sms-security, lee-20-sim-swap}. In addition, these have some usability issues such as users who lose access to their telephone numbers also lose their Signal contact lists. Finally, they hamper additional features requiring additional metadata, like user profiles.

To improve the app in terms of these shortcomings, Signal released two new features: \textit{Secure Value Recovery} (SVR)~\cite{lund-19-secure-value-recovery} and \textit{registration lock}~\cite{oleary-20-registration-lock}. Both features require the user to establish a PIN, which can be a sequence of numbers, like a traditional PIN, but also include letters and symbols. 
%SVR uses the PIN as input to a key-derivation function to derive a symmetric key.  
This key is used to recover encrypted backups of contacts and settings stored on Signal servers. The registration lock aims to prevent anyone but the original user from creating a Signal account for a phone number without the associated PIN. 

Signal's choice of naming the credential a ``PIN'' (as in, personal identification number) may not clearly indicate to the user the importance of the PIN in the Signal ecosystem. Unlike device or screen lock which is familiar to users, the in-app use of the Signal PIN is meant to achieve an app-specific purpose not satisfied by the device or operating system's features. A banking app for example might mostly be using in-app authentication to protect access to an OAuth token, while Signal has a different goal.

As Signal represents one of the first, large-scale usages of in-app PINs, in this paper we investigate to what extent do participants, both the security-/privacy-savvy and the average ones, understand the PIN feature and what effect does this have on their choice and usage? Additionally, we also investigate how participants react to Signal's PIN verification reminders that encourage users to not only select a complex PIN but regularly remind users to reenter it for verification. This feature may have been implemented because the PIN is not meant for daily use,
%used for regular functionality, 
but instead only needed in acute moments of setting up a new device with the Signal app. Finally, we examine the way participants select and compose their Signal PINs and the effect of their general understanding of the underlying Signal features to make these decisions. To this end, we consider the following research questions:
\vspace{2em}
\begin{enumerate}[leftmargin=3em, label=\textbf{RQ\arabic*}]
    \item Are participants aware of how and why in-app PINs are used in Signal?
    \label{rq1}
    \item How effective are PIN reminders in assisting participants to remember PINs?
    \label{rq2}
    \item How do participants choose and compose a PIN for Signal, and does their understanding of how these PINs are used affect that choice? 
    \label{rq3}
\end{enumerate}

We surveyed Signal users ($n=235$), asking about their understanding, usage of the Signal PIN feature, and response to Signal PIN verification. For example, we asked participants to explain the purpose of Signal PINs, in their own words. We additionally asked participants about the composition of their PIN (e.g., length, character set), if they reuse the PIN in other contexts (e.g., phone lock, in another messenger app), if they have opted out of selecting a PIN, and their response to periodic PIN verification. 

We find that only 14\,\% ($n=33$) of respondents opted out of setting a Signal PIN, and also we find a large disparity between the practices of participants who can explain the purpose of the in-app PIN authentication (who we term Signal \emph{enthusiasts}; $n=132$; 56\,\%) and those who cannot (dubbed \emph{casual} Signal users; $n = 103$; 44\,\%). 

Many enthusiasts set PINs because they thought it was required~---~initial communication from Signal indicated that it was, although it is not in current versions of the app. Many enthusiasts also specifically mentioned registration locking and cloud backups. Interestingly, when enthusiasts did not set a PIN, 44\,\% cited anti-cloud storage sentiments, indicating that they are aware of the features Signal PIN provides (e.g., cloud backups of profiles) but felt that this metadata storage did not sufficiently guard their privacy. Among casual users, 25\,\% set a PIN for generalized security reasons although they are not able to articulate those. Moreover, 13\,\% set a PIN simply because they were prompted by Signal or do not know why they actually set a PIN (16\,\%). If casual users did not set a PIN, they typically indicate that it was inconvenient (18\,\%) or they did not see the necessity (18\,\%). Their inaccurate understanding also affects this decision: 24\,\% state that they do not need an additional safeguard to secure access to their Signal app although the PIN is not used for this purpose.

Very few participants who set a PIN indicated that they had difficulty remembering their PIN; only 12\,\% said they {\em occasionally}, {\em frequently} or {\em very frequently} have difficulty remembering. When interacting with the periodic reminders to verify their PIN, 59\,\% confirm their PIN {\em frequently} or {\em very frequently}. Only 24\,\% of all participants confirm their PIN {\em rarely}, {\em very rarely}, or {\em never} when prompted, yet, here the behavior of enthusiasts and casuals diverges: 16\,\% of the latter tend to ignore the reminder prompt compared to 28\,\% of the enthusiasts. In addition, 45 or 24\,\% of the participants who currently use a PIN disabled these reminders. When asked why, 67\,\% of the enthusiasts mention that they use a password manager while casuals are mostly annoyed (42\,\%) or do not feel it is necessary to be reminded (33\,\%).

We also find that enthusiasts' PINs are more password-like, often containing numbers, letters and symbols. Compared to casuals, enthusiasts on average choose PINs with an additional 1.3 digits, 3.0 letters, and 1.3 special characters.

Moreover, many participants, particularly enthusiasts, use a password manager to store their Signal PIN, which additionally increased the complexity of their PIN: password manager users selected PINs with an additional 2.1 digits, 5.3 letters, and 3.1 special characters compared to non-password manager users. A number of participants, both enthusiasts and casuals, noted the reuse of their Signal PIN in other contexts, apps, and as their screen lock, yet, 76\,\% of the participants who use a PIN within Signal said they do not reuse it.

In short, it appears Signal's core audience of privacy-conscious enthusiasts is using the PIN effectively, however, this roll-out may have been affected by inconsistent communication. Some earlier versions of the app made PIN creation a requirement.  In addition, Signal PINs can contain letters and special characters. Weak Signal PIN choices can have consequences for those that choose secure PINs as secure communication requires both parties to be secure. We would recommend that Signal consider adding features to encourage better choices, like an improved blocklist, or even re-branding Signal PINs to more accurately depict their use, like ``Account Recovery Passwords,'' which could help users apply the right context during selection and storage of this credential. Though our focus is on Signal, our results may inform communication strategies of other app developers, since account recovery and registration lock features are common in secure messaging.

\emph{All our findings were shared with the Signal developers.}

%% file: sections/02-background.tex
%-------------------------------------------------------------------------------
\section{Background}
%-------------------------------------------------------------------------------
\label{sec:background}

Signal is an open source app and service, developed and operated by the non-profit Signal Technology Foundation. Signal implements  the underlying Signal protocol which includes forward secrecy~\cite{roesler-18-more-is-less, cohn-20-signal-analysis} and is used by other secure messaging clients, like WhatsApp~\cite{marlinspike-16-google-signal} and Facebook Messenger's secret conversation feature~\cite{marlinspike-16-facebook-signal}.
Signal boasts more privacy consciousness in its design and implementation, eschewing linkages to an identity or collection of metadata, as compared to its competitors, like Telegram, WhatsApp, or Threema 
\cite{whatsapp-21-updated-privacy-policy,singh-21-signal-brian-acton}.
Hereinafter, when we refer to \emph{Signal,} we mean the app/service and not the protocol unless otherwise specified. 

Given its focus on privacy, Signal historically relied on a user's mobile phone number as an identifier, reasoning that this system was already in place. This approach also makes migrating to a new device easier for users when using the same phone number, as long as the user's contacts were already backed up by other means. Other app settings, e.g., groups and blocked contacts, were formerly not backed up. 

Additionally, 
receiving a valid SMS with a security code was sufficient to (re-)establish an account with Signal to send/receive encrypted messages. Unfortunately, phone numbers can be subject to SIM-swapping attacks~\cite{andrews-18-your-digits, jover-20-sms-security, lee-20-sim-swap}, whereby an attacker is able to register an existing phone number with a new mobile SIM card, effectively stealing a user's account on Signal. 

To address both backing up device settings and preventing account hijacking, Signal introduced two new features: \textit{Secure Value Recovery}~\cite{lund-19-secure-value-recovery} and \textit{registration lock}~\cite{oleary-20-registration-lock}. Both services require an additional authentication check, namely a PIN, and in the rest of this section, we describe Secure Value Recovery, registration lock, and how Signal rolled out PINs.

\paragraph{Secure Value Recovery}
\label{subsec:SVR}
\emph{Secure Value Recovery} (SVR) enables encrypted backup and recovery of the Signal app settings, including contacts, profile, and group memberships. The backup data is encrypted and stored on Signal's servers. 
When a user migrates to a new device, the goal is to restore this data into the new app installation. As the decryption needs a key, the user has to choose, recall, and enter a PIN which is input to a key-derivation function. The resulting symmetric master key is used to further derive the backup encryption key.

\paragraph{Registration Lock}
\label{subsec:RL}
The registration lock is an optional feature that binds the Signal PIN to the user's phone number. This way knowledge of the PIN is required as a second authentication factor in addition to the ability to receive an SMS with a one-time security code.
This approach protects Signal from attacks like SIM swapping~\cite{andrews-18-your-digits, jover-20-sms-security, lee-20-sim-swap} where an attacker can obtain the SMS code.

To realize this functionality, the protocol uses the symmetric master key that is calculated as part of SVR, this time to derive a 32-byte registration lock hash. This value is used similarly to a password: it is sent to the server to authenticate the user.  If the calculated registration-lock hash matches the one that is stored on the Signal server, the SMS code is sent. If not, the SMS code will not be sent and the registration of the phone number cannot be completed. 

On the other hand, if an account needs to be migrated to a new device and the user does not know the PIN, setting up the account with the phone number is only possible after 7 days of inactivity. After this time span, the server's registration lock hash (of the PIN) expires and a new account can be created. However, the counter will be reset each time the client connects to the Signal server which happens when receiving or sending messages. Additionally, the iOS or Android apps make requests on a regular basis to keep the PIN hash alive even if the app itself is used infrequently.
 
\paragraph{Signal PINs}

Unlike PINs used to authenticate to gain access, e.g., unlocking your phone, the Signal PIN is used as a secondary authentication factor when moving an account from one device to another. A user does not need to enter the PIN to use Signal once it is installed on a particular device. However, Signal has a separate setting that locks the application from unauthorized access by forcing the user to verify their mobile phone's unlock authentication, e.g., the PIN used to unlock the device. 

Also different than unlock authentication PINs, if a user forgets their Signal PIN while maintaining access to the Signal app, it can be reset without any repercussions as the current secure messaging keys can serve the purpose of authentication. After resetting the PIN, the SVR-encrypted backup can be re-encrypted and uploaded to Signal's servers, and the registration lock hash can be regenerated.  

Communicating the purpose of the PIN to users, including all the features it does and does not support, is not a straightforward task. While Signal published an article explaining the technical details of SVR and registration lock \cite{oleary-20-registration-lock}, explaining it to all users remains a challenging task. Signal also originally required a user to establish a PIN, but later made that choice optional.  

Finally, as the Signal PIN is only needed at acute moments, Signal employs periodic PIN reminders to help users memorize their PIN.  These reminders to verify a PIN are spaced at regular intervals, starting at 12 hours, then 1 day, 3 days, 7 days, and every 14 days.  Figure~\ref{fig:verify} shows the prompt that is shown to users for this purpose.

%% file: sections/03-related-work.tex
%-------------------------------------------------------------------------------
\section{Related Work}
%-------------------------------------------------------------------------------
\label{sec:previousWork}
%-------------------------------------------------------------------------------

The Signal PIN is used for authentication in the mobile setting, an area that has received a good deal of attention in the literature.  For example, password usability on smartphones is studied along with shoulder-surfing resistance by Schaub et al.~\cite{schaub-12-passwords-on-smartphones}.  Aviv et al. consider the advantage gained by a guessing attacker as a result of screen smudges~\cite{aviv-10-smudge}.  User choices for mobile unlock methods are investigated by Harbach et al.~\cite{harbach-14-hard-lock-life}, who find that alphanumeric PINs are less popular than numeric PINs.  

In our study, we note a number of casual users with limited comprehension, a theme also observed in other circumstances of secure messaging. Abu-Salma et al.~\cite{abu-salma-17-secure-messaging} noted that security and privacy is not always a leading driver in the adoption of a secure messenger like Signal, but rather community pressure of wanting to be able to reach specific contacts. De Luca et al.~\cite{de-luca-16-attitudes-towards-messaging} and Das et al.~\cite{das-14-social-influence, das-14-social-proof, das-15-social-influence} come to a  similar conclusion and show that the influence of social factors is not only limited to the adoption of messengers but security tools in general. Abu-Salma et al.~\cite{abu-salma-17-secure-messaging} further note that many users have misconceptions about the security of messaging, e.g., they perceive SMS as secure for sensitive communication. Oesch et al. conduct a user study confirming user misconceptions and finding that group-chat users tend to manage security and privacy risks using non-technical means such as self-censorship and manually inspecting group membership~\cite{oesch-20-understanding}.

%Signal
In general, Signal and other secure messaging services often face the problem of explaining secure protocols, however, authentication ceremonies are challenging for users to understand~\cite{vaziripour2017you,vaziripour2018action}. To address this issue, Wu et al. offered a redesign of the authentication ceremony that emphasizes comprehension~\cite{wu19-something-isnt-secure}. Vaziripour et al.~\cite{vaziripour-19-social-media-ceremony}, on the other hand, suggested to partially automate the ceremony by using social media accounts. The Signal PIN is used for key derivation and is an example of a \emph{usable encryption} scheme in the real world.  These have been previously studied by Ruoti et al.~\cite{ruoti-13-confused-johnny} who propose a secure email system and study varying levels of user transparency and automation. While Signal  aims for automatic key management and automatic encryption, Ruoti et al. find that users had more trust in an approach that emphasized manual steps and therefore comprehension. 
While our research aims to understand how comprehension affects users' PIN practices, similar efforts to better communicate about this feature would likely help users. 
As a user chosen secret, Signal PINs also relate to the choice of traditional PINs. Initial research on user choice of numeric PINs was done by Bonneau et al.~\cite{bonneau-12-pin} who found that dates are particularly prominent. Kim et al. also found dates to be common, as well as digits in sequence, like \emph{1234}~\cite{kim-12-pin-policies}. Signal's PIN in fact takes this advice and blocks digits in sequence.
Wang et al.~\cite{wang-17-pin} derived numeric sequences from leaked password datasets, and Bonneau et al.~\cite{bonneau-12-entropy} measured their guessability. Wang et al. found that PINs generally are easily guessable in online attacks (where an attacker only has a limited number of attempts or is rate-limited), and surprisingly 6-digit PINs more so than 4-digit PINs. This line of research was confirmed and extended in the context of mobile unlock PINs by Markert et al.~\cite{markert-20-pin-blocklist}. Markert et al. also showed that a well-sized blocklist of PINs, when enforced, can significantly improve PIN-guessing resistance in an online setting.

Recently, Khan et al.~\cite{khan-20-widely-used} and Casimiro et al.~\cite{casimiro-20-create-reuse-pins} studied PIN reuse across different contexts. Both find that reuse is rampant, and that users tend to have a small set of PINs they use regularly. In our work we also find that certain kinds of PIN reuse is common for Signal PINs, such as for an ATM/Credit/Payment card.
As Signal PINs are generally chosen and entered on mobile devices, users may be less inclined to choose hard-to-guess, full-fledged, alphanumeric passwords with special symbols. (Recall that a Signal PIN can have numbers, letters, and special symbols.) Melicher et al. studied user selection of passwords on mobile devices~\cite{melicher-16-smartphone-passwords}, finding that the limitations of the keyboard setting may lead to more easily guessable and weaker passwords. 

In our work, we find that participants using a password manager are more likely to select strong Signal PINs. Unfortunately, in the mobile setting, users remain challenged in using password managers. Seiler-Hwag et al. investigated common password managers on smartphones~\cite{seiler2019don}, finding that all score poorly on standard usability metrics. Even when a password manager is adopted, using the password generation feature is not a given for all users.  Pearman et al.~\cite{pearman-19-use-pw-manager} studied why users do (and do not) adopt a password manager and find that even those that do use a password manager may not use the password generation feature.

To the credit of the Signal team, they understood that the Signal PIN is unlike the case of mobile unlock authentication where a typical user unlocks the device multiple times per day. Instead, they realized an infrequently-used PIN is much more subject to being forgotten by the user. So they employ the well-known technique of \emph{graduated interval recall} (also called \emph{spaced repetition}).  While the positive effects on recall rates have been shown in multiple studies~\cite{spitzer-39-studies-in-retention,pimsleur-67-memory-schedule, melton-70-the-situation, landauer-78-optimum-rehearsal}, including the memorability of passwords~\cite{horcher-09-building-a-better, mujye-13-complex-passwords, bonneau-14-56bit, blocki-15-pao, novak-17-modeling-security,schechter-15-learning-assigned-secrets}, the usage of it in this context is novel. The deployment of Signal's periodic reminders to verify the PIN offers a real world example of the effectiveness of this strategy.

%% file: sections/04-method.tex
\section{Method}
%-------------------------------------------------------------------------------
\label{sec:method}
We conducted a user study of $n=235$ Signal users recruited to complete a survey about their understanding and strategies for managing their Signal PINs. In this section, we provide details of the survey, recruitment, limitations, and ethics. 

%-------------------------------------------------------------------------------
\subsection{Study Design}
%-------------------------------------------------------------------------------
\label{sec:userStudy}

We recruited participants in two samples. The first sample was from Reddit, the Signal Community Forum, and snowballing; the second sample via Prolific. For participants completing the study on Prolific, we first used Prolific's built-in screening to only recruit participants who use Signal, and as this pool was still insufficient, we used a single screener question (Appendix~\ref{app:screen}) as part of a two-part recruitment, where participants noted which messaging app they used. Those using Signal were invited to the main study. The entire survey is provided in Appendix~\ref{app:survey}, and it took participants 7 minutes, on average, to complete. 

\begin{enumerate}[noitemsep]

\item {\em Informed Consent}: All participants were informed of the procedures
  of the survey and provided consent. The informed consent notified
  participants that they would be asked to complete a short survey that asks questions about how they select PINs and how they feel about Signal's implementation. 
  
\item {\em Signal Usage}: Participants must indicate they are a Signal user answering the question: ``Do you use Signal?''(\ref{app:survey:q0}) All participants who responded in the affirmative continued with the survey. 

\item {\em PIN Comprehension and Usage}: Participants were now prompted with the text: ``PINs are a new feature provided by Signal. In your own words, please explain how PINs are used by Signal,'' (\ref{app:survey:q3}) followed by ``Did you set a Signal PIN?''(\ref{app:survey:q4}) and why they did (\ref{app:survey:q5a}) or why they did not (\ref{app:survey:q5b}). Those who did not set a PIN skipped ahead to \ref{app:survey:q25}.

Those who did set a PIN were asked if the PIN was since disabled (\ref{app:survey:q6}), and if so, why (\ref{app:survey:q7}). We also asked participants who still had their PIN enabled if they have difficulty remembering their PIN (\ref{app:survey:q8}), and what they would do if they forgot their PIN (\ref{app:survey:q9}).

\item {\em PIN Reminders}: We then asked a series of questions (\ref{app:survey:q10}-\ref{app:survey:q14}) on Signal's periodic PIN reminders (cf. Figure~\ref{fig:verify}), including if participants currently have the reminder set; for those who do, how frequently they verify the PIN when prompted; and if they disabled it, why.

\figverifypin{}
    
\item {\em PIN Reuse and Sharing}: Participants were asked to report if they reuse their Signal PIN in other contexts, such as mobile device unlock (\ref{app:survey:q15}), ATM and other payment cards (\ref{app:survey:q16}), and other mobile applications (\ref{app:survey:q17}).  In addition, participants were also asked if they have shared their PIN with friends or family (\ref{app:survey:q18}). These questions were derived from related work on PIN usage~\cite{casimiro-20-create-reuse-pins, khan-20-widely-used}. 

\item {\em PIN Selection and Composition}: The survey continued with a series of questions about PIN length and composition, as well as the perceived strength of the PIN  (\ref{app:survey:q19}-\ref{app:survey:q24}).

\item {\em Other Messengers}: The survey continued by asking about the use of PINs in other messengers, including Facebook, Skype, Telegram, WeChat, and WhatsApp (\ref{app:survey:q29}). We also asked if the Signal PIN is reused in any other messenger as well as the reasons for doing or not doing so (\ref{app:survey:q30a}/\ref{app:survey:q30b}).
    
\item {\em Demographics:} Finally, we asked about demographics (\ref{app:survey:d1}-\ref{app:survey:d5}), including age, gender, and IT background. 

\end{enumerate}

%-------------------------------------------------------------------------------
\subsection{Recruitment \& Demographics}
%-------------------------------------------------------------------------------
\label{sec:recruitment}
We recruited a total of $n = 235$ participants. 
Of those 170 were recruited from Reddit, the Signal Community Forum, and snowballing, and 69 were recruited on Prolific. We posted to Reddit's r/SampleSize and r/Signal forums; and the Signal Community Forum. We decided against a fixed payment for these participants in favor of not collecting any personally identifiable information, e.g., an email address to offer a gift-certificate via a raffle, and thus these participants took the survey voluntarily without compensation. 

We used Prolific's built-in custom prescreening filters, which allow researchers to post a study to participants that meet specific criteria, e.g., residing within the US. We applied the custom prescreening for Prolific members who indicated Signal is one of the ``chat apps'' they use regularly. We were able to recruit 69 participants this way, each paid GBP\,1.50. To expand the Prolific pool, we also employed a custom screening survey to find other Signal users, recruiting 500 responses (paying GBP\,0.15). Those who indicated that they used Signal were invited to the main study (paying GBP\,1.50). We were able to recruit an additional 11 participants this way. 

\tableDemo{}

As shown in Table \ref{tab:demo}, the demographics of our sample is skewed toward a younger, more male-identifying, and more IT-oriented group. On the other hand, our participants reside in many different countries increasing the generality of our results.
Of the 235 participants, Germany accounted for (68; $29\,\%$), the USA for (61; $26\,\%$); the UK for (24; $10\,\%$). The rest of the world was the largest group with (82; $35\,\%$). The actual demographics of the Signal community at large are unknown, so the skew towards a certain participant pool may reflect our recruiting strategy or may be influenced by the makeup of the underlying community.  We observe that at $75\,\%$, males make up the largest cohort.  Similarly, at $64\,\%$, those with IT-focused education or employment make up a majority of participants.  In terms of education, bachelor's and master's groups combined account for $55\,\%$ of participants.  Our group of enthusiasts is also male-dominated: self-identified males outnumber females more than 8:1.  
Finally, we note that among enthusiasts, the IT-focused group is substantially larger at 3:1, while the figures are more balanced for casuals: about 1.2:1.  It is reasonable to surmise that an IT background makes one more likely to be a enthusiast --- put another way, Signal's existing communication strategy about the Signal PIN appears to be more effective for those with an IT background.

%-------------------------------------------------------------------------------
\subsection{Limitations}
%-------------------------------------------------------------------------------
\label{sec:limitations}
As this study took place online, it shares the usual limitations of many online studies, such as finding a representative recruitment.  On the one hand, our sample may not be a representative sample of all Signal users. Though we did not explicitly sample enthusiasts and casuals separately, we found that comparatively more enthusiasts were recruited via Reddit and Signal Community Forum, which led us to perform additional sampling from Prolific.  

As an online survey, this study necessarily relies on self-reported data.  With regard to security and privacy user studies, Redmiles et al.~\cite{redmiles-19-generalize} show online-survey responses generalize quite readily to the broader population.  Additionally, we conducted extensive pilot testing among members of our research groups and trusted colleagues to identify any ambiguities in our survey questions.

Another limitation is that participants' responses may suffer from the well-known tendency toward providing socially-desirable answers~\cite{macoby-54-social-desirability, fisher-93-social-desirability}. For example, it is possible that PIN reuse is more prevalent than our study suggests, or that people choose PINs that are shorter and have less-diverse composition. The same holds for questions where we asked participants about their own understanding, where they might have looked up answers on Signal's website. Despite this possibility, the answers provided appeared unique and participants provided many apt phrases to describe the situation. Additionally, we did not find responses that were directly cut and paste from Signal's website.

%-------------------------------------------------------------------------------
\subsection{Ethics}
%-------------------------------------------------------------------------------
\label{sec:ethics}
The study was administered at an institution that does not have an Institutional Review Board (IRB), but we still followed all appropriate study procedures similar to studies that obtained IRB approval. For example, participants were informed about the nature of the study, participated voluntarily, and could opt-out at any time. Additionally, we conformed with the ethical principles laid out in the Menlo Report~\cite{dhs-12-menlo-report}, e.g., we minimized any potential harm by not collecting any personally-identifiable information from our participants.

As described above, we completed two recruitments, one with paid and one with unpaid participants. Unpaid participants were recruited via Reddit, Signal Community Forum, and snowballing. We decided not to pay those participants as paying a comparatively small amount did not appear to withstand the harm that went along with collecting email addresses. Additionally, as this community tends to be more privacy-conscious, doing so might have depressed participation. Participants recruited via Prolific were paid GBP\,1.50 for successfully completing the main survey, as this amount is in line with the recommended rewards on Prolific \cite{prolific-18-reward}. 

%% file: sections/05-results.tex
%-------------------------------------------------------------------------------
\section{Results}
%-------------------------------------------------------------------------------
\label{sec:results}
\vspace{.25em}
In this section, we present the results of our study of $n=235$ Signal users. For the structure of the section, we follow our three research questions: we start by analyzing the comprehension of the usage of PINs in Signal (\ref{rq1}), continue with user responses to the reminder feature (\ref{rq2}), and conclude with PIN selection and composition (\ref{rq3}). 

For qualitative analysis, we had a primary coder code all the qualitative responses, producing an initial codebook. A secondary coder used that codebook to independently code the same responses, and afterward, the two coders met to resolve differences to produce a final codebook. The primary coder then used that final codebook to re-code the data. The codebook used for each qualitative question can found in Appendix \ref{app:codebook}, Tables~\ref{tab:codebook-q3}--\ref{tab:codebook-q30}.

\subsection{RQ1: Comprehension}
\label{sec:comprehension}

As part of \ref{rq1}, we seek to understand Signal users' awareness and understanding of PINs and how they fit into the Signal ecosystem. To answer this question, we divide the participant pool by those that have or have not adopted a Signal PIN, and also by those that demonstrate understanding of how Signal uses the PIN.

\paragraph{Understanding Signal PINs} 

After indicating if they are a Signal user (\ref{app:survey:q0}--\ref{app:survey:q2}), we first ask participants to describe how Signal PINs are used in their own words (\ref{app:survey:q3}): {\em PINs are a new feature provided by Signal. In your own words, please explain how PINs are used by Signal.} These responses were coded by comprehension and accuracy; specifically, we seek to understand if the participants recognized that PINs are used for SVR and registration lock. Participants who accurately described the usage of Signal PINs were coded as {\em enthusiasts} ($n=132$; 56\,\%), and those who could not describe Signal PIN usage were coded as {\em casual} Signal users ($n=103$; 44\,\%).

\figUserbreakdownBar{}

We observed many different ways of capturing the main elements of how PINs are used by Signal. Many of the enthusiasts were even able to demonstrate a deep understanding, for example P10 said:
\begin{quote}
    \small \em
    ``It protects data like settings and group membership and signal [sic] contacts that will be stored on Signal's servers using SVR. Previously this was only stored locally on a user's device and was lost upon device reset or getting a new device unless a full backup was made on Android.''
\end{quote}

Participant responses were assigned one or more codes based on the aspects correctly described.  
Overall among enthusiasts, the most popular codes were backup (65; 49\,\%), encryption (45; 34\,\%), contacts (31; 24\,\%), and registration (23; 17\,\%). Some also noted settings (8; 6\,\%), profile (4; 3\,\%) or groups (3; 2\,\%), which are also secured via a Signal PIN during backup, and a few specified key derivation (7; 5\,\%). Some also mentioned that PINs were part of a process for Signal to move away from using phone numbers for identity (6; 5\,\%). A handful of enthusiasts also expressed anti-cloud sentiments when asked about Signal PINs (2; 2\,\%), suggesting that they understood that the PINs play a role in the encrypted cloud backup functionality of SVR, and that they are opposed to that design direction. 

For the casual users, a majority (57; 55\,\%) provided non-answers, or answers that do not indicate any understanding of the way the Signal PIN is used. The answer of P47 accurately summarized the reasoning we observed for many casual users: 
\begin{quote}
    \small \em
``I don't understand their purpose very well. I thought that they might be using the PIN system to verify the identity of the person using signal (if for instance someone unauthorized gained access to the phone), but the way that pin entry is optionally offered every few weeks doesn't align with such a purpose. as such, I have no idea what they're trying to accomplish.'' 
\end{quote}

As the majority of casual users didn't know or provided non-answers, there are many other examples to choose from, including ``I initially thought it was used as a local PIN to unlock the app on my phone. It doesn’t do that so I have no idea how it works,'' from P62.
Additionally, many casual users falsely associated PINs with securing messages (21; 20\,\%) although messages are not part of the backed-up data and are not protected by the PIN, as explained by P183: ``Keep your messages on Signal encrypted via use of the PIN.'' 

An equal number felt that the PIN locks the Signal app (21; 20\,\%), while in fact that functionality is called Signal Screen Lock and is not related to the Signal PIN~---~for that feature, Signal simply re-uses the device's existing PIN, biometric, or other authentication scheme. An example of this response is from P37: ``Protect application from opening from an unlocked phone.'' Similar responses show this is a common misconception: ``Pins are used to prevent unauthorized access to the app'' from P227.  Some individual participants also mentioned security as a general topic, without further describing it (2; 2\,\%), or associated the PIN with inconvenience (1; 1\,\%).

\paragraph{Why did participants set a PIN?}
In addition to knowing if participants understand the usage of the PIN, we also want to analyze how many actually set a PIN in their Signal app. In total, 202 or 86\,\% of all 235 participants adopted a PIN. If we further divide those 202 participants based on their understanding, we see that more enthusiasts (116; 57\,\%) than casuals (86; 43\,\%) set a PIN.

To get a deeper understanding, \ref{app:survey:q5a} asked participants to explain their decision. By far the most popular reason, equally distributed among enthusiasts and casuals, is \emph{security}: 48 or 24\,\% mentioned it in their answer. Once again, we find that enthusiasts display a detailed, in-depth understanding, exemplified by P14: 

\begin{quote}
    \small \em
``I want to be able to use secondary identifier once it becomes available and not to lose my contacts that are not in my phone's contacts list. I also want to be secure against SIM-swap attacks.''
\end{quote}

This code is followed by participants mentioning that they were required to set a PIN (33; 16\,\%).  Among enthusiasts, we observed 25 that mentioned it was required (or $22\,\%$). P164 said  ``I had absolutely no choice if I wanted to continue to use Signal. Eventually, the box asking you to create a PIN kept you from opening any of your messages until you did what it wanted.'' 

This response may reflect the changing nature of the PIN requirement. Initially, it was required and then in a subsequent version, merely encouraged. The enthusiast-casual split here suggests perhaps more enthusiasts were early adopters of the Signal PIN.  Another theme, of setting a PIN due to annoyance (12; $11\,\%$) may also reflect this changing communication strategy for Signal PINs. See for example Figure \ref{fig:create:v1}, showing the initial prompt used by Signal to ask users to create a PIN; the prompt has subsequently been updated to \ref{fig:create:v3}, current as of this writing. Observe the communication is also different when a user wishes to change their PIN as shown in Figure~\ref{fig:create:v2}, again current as of this writing.

Enthusiasts also regularly noted registration lock as a reason to set a PIN (14; $12\,\%$). P3 said ``The PIN stop [sic] others from registering as me, and also protects access to my account details (profile, settings, contacts) if my device is misplaced.''

Casual Signal users noted \emph{security} most frequently (26; $29\,\%$), but did so in a more general way as seen in this quote from P141: ``for security and for reassurance if device gets stolen.'' Additional codes include \emph{don't know} (16; $18\,\%$) and \emph{prompted} (13; $15\,\%$), suggesting that many casual users selected a PIN simply because they were prompted to do so and had no other underlying motivations. For example, P155 responded ``I trusted the app and just did it when prompted.''  

\paragraph{Why did participants not set a PIN?}
A total of 33 (14\,\%) participants chose not to set a PIN (see Figure~\ref{fig:userbreakdown-barNew}). A roughly equal number of enthusiasts and casual Signal users did not set a PIN: 16 enthusiasts ($12\,\%$) did not set a PIN and 17 (17\,\%) casual Signal users did not set a PIN. A $\chi^2$ test revealed no significant differences between the groups.

When these $n=33$ participants described why they did not set a PIN (\ref{app:survey:q5b}), there were a number of differences. Both casual (3; 18\,\%) and enthusiasts (4; 25\,\%) described PINs as inconvenient, but casual users were more likely to note that either they do not need a Signal PIN (3; 18\,\%) or that their phone lock provided security (4; 24\%). For example P227 noted that their \enquote{\ldots phone is always locked} and \enquote{Additional authentication seems unnecessary.}

Enthusiasts expressed distrust as a reason for not setting a PIN. Either this distrust is in the security of PINs for key derivation and management (3; 19\,\%), or they distrust cloud storage (7; 44\,\%). Distrust of cloud storage stems from privacy concerns with the SVR feature that backs up contacts and settings. P216, for example stated, that they \enquote{had no desire to have any contact data uploaded,} and  P207 said \enquote{i [sic] do not want to store personal information in the cloud.}

\figUserbreakdownBarNew{}

\paragraph{Why do participants disable PINs?}
On top of the 33 users who declined to set a PIN, a total of 11 (5\,\%) set a PIN and then later disabled it: 5 (45\,\%) enthusiasts and 6 (55\,\%) casual users, as shown in Figure~\ref{fig:userbreakdown-barNew}. When asked to explain  why they disabled their PIN (\ref{app:survey:q7}), participants mentioned that the PINs were annoying (4; 36\,\%) or inconvenient (2; 18\,\%), which may be related to the periodic verification reminders. P212 explicitly mentioned the \enquote{verification overhead.} Anti-cloud hesitation to store data on Signal's servers led (3; 27\,\%) participants to disable their PIN: \enquote{Don't want my data stored on their server} (P193). We also observed (2; 18\,\%) participants who simply stated that they \enquote{do not need it} (P206). 

\subparagraph{\ref{rq1} Results Summary}
Signal users in our sample break down into two groups: enthusiasts who were aware of the features Signal PINs enabled, and more casual Signal users who were unable to describe how PINs are used within Signal. In both groups, though, setting a Signal PIN was highly prevalent. Only 33 of the 235 respondents chose not to set a PIN. Among enthusiasts, their choice to not set a PIN stemmed from either distrust in the key-derivation process or hesitancy to store information in the cloud generally. Casual users did not set a PIN because of inconvenience or a false belief that other authentication mechanisms, like locking their phone, provided adequate protection. When participants disabled their Signal PIN, inconvenience or annoyance were often cited, sometimes referring specifically to the periodic reminders. 

\subsection{RQ2: PIN Recall and Reminders}  
In this section, to address \ref{rq2}, we consider how participants remember their PINs and their reactions to the periodic PIN verification reminders. Throughout this section we consider the $n=191$ participants who still have their PIN enabled, and not the 11 participants who since disabled their PIN.

\paragraph{Forgetting PINs}
We asked the ($n=191$) participants who still use a Signal PIN in \ref{app:survey:q8} if they encountered difficulty in remembering their PIN. Overwhelmingly, $89\,\%$ of participants  ($n=170$) indicated that they {\em never}, {\em very rarely}, or {\em rarely} have difficulty remembering their PIN (see Figure~\ref{fig:PINMemorabilityAndVerification}; {\em top}). We compared the response to this question from enthusiasts ($n=106$) and casual ($n=85$) Signal users who still had their PIN enabled, and we found no statistical differences.

We asked participants in \ref{app:survey:q9} what they would do if they forgot their Signal PIN. (Note that the PIN is not required to use Signal for messaging, and can be reset at any time in the settings menu.) Many enthusiasts noted that their PIN was stored in their password manager (45; $42\,\%$), and they would simply look it up. Fewer casual participants mentioned a password manager (12; $15\,\%$). A number of participants did not know what to do (27; $25\,\%$ enthusiasts and 33; $40\,\%$ casuals), while a few casuals suggested they would contact Signal (4; $5\,\%$) and two enthusiasts said they would reinstall the app (2; $2\,\%$). Others believed that their Signal account is now unrecoverable (2; $2\,\%$ enthusiasts and 3; $4\,\%$ casuals);
some would create a new account (4; $4\,\%$ enthusiasts and 5; $6\,\%$) casuals. A handful (2; $2\,\%$ enthusiasts and 4; $5\,\%$ casuals) denied that they would forget stating  \enquote{It is a PIN I use for my bank cards} (P145), for example.
%, or that it was written down \reminder{($n=X$;Y\,\%)}. 
A small number of participants noted that they would wait 
(8; $7\,\%$ enthusiasts and 4; $5\,\%$ casuals), aware that the registration lock expires after 7 days of inactivity.

\figLikert{}

\paragraph{Periodic Verification}
Perhaps recognizing that Signal PINs are only truly required when transferring a Signal account to a new device, Signal decided to employ \emph{graduated interval recall}~\cite{pimsleur-67-memory-schedule} (or, \emph{spaced repetition}) that regularly prompted participants to verify their PIN when opening the Signal app. An example of such a reminder is found in Figure~\ref{fig:verify}. To our knowledge, Signal is the first mainstream app to implement such a feature.

We first asked participants if they were aware of the PIN verification reminders (\ref{app:survey:q10}). Most participants ($n=176$; 92\,\%) indicated that they were aware, and a follow up question (\ref{app:survey:q12}) asked if they have since disabled the reminders. Seventy-four percent ($n=131$) of participants have the periodic PIN verification enabled, and many still verify their PIN when prompted. Seventy-six percent ($n=135$) of participants either {\em occasionally}, {\em frequently}, or {\em very frequently} verify their PIN when prompted. When dividing this data by enthusiasts and casual Signal users (see Figure~\ref{fig:PINMemorabilityAndVerification}; {\em bottom}), we did not observe significant differences between frequency of PIN verification using a Mann-Whitney U test.   

The remaining 23\,\% ($n=45$) disabled the PIN reminders. These 45 participants were asked why they disabled the reminders (\ref{app:survey:q14}): (23; 51\,\%) mentioned doing so because they use a password manager. P63 said ``I don't remember my PIN, it's stored in my password manager, frankly, I don't even want to remember it.''  Ten ($22\,\%$) said there was no need or their PIN was already memorized, and a further (11; 24\,\%) found the reminders annoying. These figures suggest that the periodic reminders are generally viewed as beneficial, or at least not substantially invasive enough to warrant disabling them. As we rely on self-reported data, we do not independently verify PIN recall rates.

\paragraph{Password Manager Usage}
We found a large amount of password manager (PM) usage in our study. These reports were entirely unprompted as PMs were not mentioned in any survey material. Thirty-one percent ($n=62$) indicated that they use a PM in response to questions regarding either what they would do if they forget their PIN \ref{app:survey:q9} or how they select their PIN \ref{app:survey:q20}. As we did not explicitly ask about PM usage, the true number of PM users might be higher.

More striking is the combination of the classification of enthusiasts and casual participants combined with that of PMs: (52; 83\,\%) of the 62 participants who said they use a PM were enthusiasts. Or, 50\,\% of the 103 enthusiasts who have a PIN enabled use a PM. Only (10; 14\,\%) of the 73 casual Signal users using a PIN mentioned PMs as a mechanism to either select or recall their PIN. Put another way, participants who mentioned a PM were overwhelmingly enthusiasts.

\subparagraph{\ref{rq2} Results Summary} 
Participants indicated that they have little difficulty remembering their PIN, many stating that this is a PIN they use all the time and thus would {\em never} forget it. A large number of participants, notably half of PIN-using enthusiasts, use password managers to both select and recall their Signal PIN, and are thus, not concerned with forgetting their PIN. Reactions to Signal's periodic PIN verification requests were more mixed, but overwhelmingly participants verified their PIN when prompted. Roughly a quarter of participants disabled periodic PIN verification; most did so because they use a password manager. Others stated that the PIN was already memorized, so there was no need for the reminders, and some simply found the reminders annoying. Overall, since 76\,\% of participants reported verifying their PIN when prompted, we conclude graduated interval recall used for Signal PIN verification is generally embraced by users, though the effectiveness of this intervention is obviously an area that deserves future work.

\figReuse{}

\subsection{RQ3: PIN Reuse and Composition}
In this section, we explore selection strategies of Signal PINs by asking participants if they reuse their Signal PIN in other contexts; the composition of their Signal PIN with respect to numbers, digits, and special symbols; and the perceived security of their Signal PIN in comparison to other PINs they use. 

\paragraph{PIN Reuse}
To explore the many ways in which PINs are reused, we adopted questions from Khan et al.~\cite{khan-20-widely-used} and Casimiro et al.~\cite{casimiro-20-create-reuse-pins} regarding PIN usage, more broadly. The responses of $n=191$ participants using a Signal PIN are found in Figure~\ref{fig:reuse}, broken down by enthusiasts and casual users.

First, as a mobile application, we asked participants if they used their smartphone unlock PIN as their Signal PIN (\ref{app:survey:q15}). Thirteen percent ($n=26$) did so, composed of 12 enthusiasts and 14 casual users. In \ref{app:survey:q16}, we asked if they used the Signal PIN in other contexts, ranging from ATM/Credit/Payment cards, to garage door codes, gaming consoles, and voice mail. (Refer to Appendix~\ref{app:survey} for the full list, derived from Khan et al. and Casimiro et al.) Twenty-eight percent ($n=53$) of participants use their PIN in another context, consisting of (25; 43\,\%) enthusiasts and (28; 53\,\%) casual users. Among those who reused, casual users did so more often: 1.39 times on average, compared to enthusiasts who did so 1.24 times.  The most common context of PIN reuse overall was for  ATM/credit/payment cards where (17; 32\,\%) of 53 participants reused a PIN. Participants also mentioned laptop/PC authentication (13; 24\,\%) and other online accounts (11; 21\,\%). 

We also asked if participants reuse PINs in other mobile applications (\ref{app:survey:q17}): (21; 11\,\%) reported they did, and of those, 12 were enthusiasts and 9 were casual users. Most commonly, the other app was WhatsApp ($n=6$); WhatsApp implements the Signal Protocol. Other common mobile apps where this PIN was reused were banking apps ($n=5$). In \ref{app:survey:q25}--\ref{app:survey:q28}, we asked participants if they use other messenger services, such as Facebook messenger, Telegram, and WhatsApp: (183; 95\,\%) did. We also asked if they set a PIN in these services and found (49; 26\,\%) did. 

Finally, we asked if participants share their PIN with friends and family: this was rare. Only 3 participants did so, suggesting that  PINs selected for Signal are not widely shared with others and are considered confidential. 

\figStrategy{}

\paragraph{PIN Composition}
A participant's understanding of the Signal PIN's functionality had a large effect on the composition of their PIN. We asked participants what was their primary PIN selection strategy in \ref{app:survey:q20}: code frequencies summarized in Figure~\ref{fig:strategy} (with full details in Table~\ref{tab:codebook-q20} in Appendix~\ref{app:codebook}). 

Among enthusiasts, password managers (PM) were mentioned frequently (28; 26\,\%). For example  P100 noted that their ``password safe generated it.''  Some participants mentioned the name of their password manager explicitly, like KeePass or Bitwarden.  Far fewer casual Signal users (6; 7\,\%) mentioned a PM.  The most-frequent code among casuals was \emph{memorable}: (30; 36\,\%), choosing a PIN easy to remember; among enthusiasts it was second-most frequent (23; 21\,\%). For example, P7 noted their PIN was ``Complicated enough but can still be remembered.''  This result suggests that despite the prevalence of randomized password generation, most participants want to select a PIN they can remember and recall easily, rather than having to look it up in a PM. 

Interestingly, while the study of Markert et al. found dates to be the most-popular strategy for selecting a PIN, only 3 of our participants mentioned dates (2 enthusiasts and 1 casual)~\cite{markert-20-pin-blocklist}. In the study of Markert et al. with ($n=200$), \emph{memorable} was the second-most frequent code (37; 19\,\%).

We then asked participants why they select a PIN with the current ``security level'' (\ref{app:survey:q22}). (Results summarized in Figure~\ref{fig:securitylevel}; full details in Table~\ref{tab:codebook-q22} in Appendix~\ref{app:codebook}.) Among both enthusiasts (25; 23\,\%) and casual Signal (20; 24\,\%) users, many mentioned security;  P44, an enthusiast, said ``I am fairly security conscious.'' 

Casuals and enthusiasts roughly equally mentioned that they chose something that was simply \emph{good enough}: (16; 15\,\%) and (15; 18\,\%) respectively. Slightly more casual users mentioned memorability: (12; 11\,\%) enthusiasts and (18; 22\,\%) casual users.  A similar number of enthusiasts (11; 10\,\%) and casuals (6; 7\,\%) mentioned that they try to be consistent in their security choices around PINs (and authentication generally), for example ``Because I always choose this security level'' (P109).

\figSecurityLevel{}

Recall that while Signal refers to this secret as a PIN, it is not a traditional {\em personal identification number}, but rather has more of the properties of a password. We asked participants to provide metrics for how many numbers, characters, and special symbols they use in their Signal PIN (\ref{app:survey:q24}). Participants were presented a slider for each class from 0 to 12.  While it is of course possible that a participant might have more than 12 digits, as a practical matter more than this simply indicates the use of a PM, which we can see in our data. 
Results are shown in Table \ref{tab:PINComposition}.

\tablePINComposition{}

Enthusiasts on average chose PINs with an additional 1.3 digits, 3.0 letters, and 1.3 special characters, and length increased overall by 5.5 characters. Except for the number of special characters, we were able to observe significant differences between the enthusiasts and the casuals using a $t$-test with Bonferroni-correction (for 8 overlapping hypotheses).

When dividing the population by their use of PMs, the difference is even greater. (Note that more enthusiasts employed a PM.) PM users chose PINs with an additional 2.1 digits, 5.3 letters, and 3.1 special characters. Overall, they used PINs which are 10.5 characters longer on average. Using a $t$-test with Bonferroni-correction, we were able to observe significant differences for all those statistics.

\subparagraph{\ref{rq3} Results Summary}
Many participants reuse Signal PINs in a number of ways. Roughly 15\,\% indicated that they use their Signal PIN as their screen lock PIN, used to unlock their smartphone. Nearly 30\,\% noted that the same PIN is used in other contexts, most commonly as an ATM/banking/payment card PIN. The Signal PIN is also reused in other mobile apps, such as a WhatsApp PIN, serving the same purpose as a Signal PIN for SVR and registration lock. When selecting a PIN, understanding of the purpose of Signal PINs led to much more diverse PINs, both in terms of the PIN length but also the presence of special characters and symbols. Among enthusiasts, the use of a password manager was particularly prominent when selecting a PIN, as compared to more casual users. But by far the largest factor in PIN selection overall is a desire for choosing a memorable PIN. 

%% file: sections/06-discussion.tex
%-------------------------------------------------------------------------------
\section{Discussion}
%-------------------------------------------------------------------------------
\label{sec:discussion}

\paragraph{Communicating about Signal PINs}

Our data show Signal's communication about the PIN feature has been effective for its traditional community of privacy enthusiasts. Without prompting, participants told us they learned about the PIN by reading blog posts, the Signal website, and tweets. Casual users, on the other hand, were much less likely to have exposure to these other sources. For this reason, in-app or in-the-moment resources nudging casual users in a more secure direction would almost certainly be of benefit.  

As explained in Section~\ref{sec:background}, the case of Signal is especially challenging.  While users are surely familiar with PINs as used in smartphone-unlock and payment-card scenarios, Signal PINs are actually used to \emph{infrequently} derive encryption keys for SVR and \emph{infrequently} act as a password for registration lock. Yet, despite the text in the Signal PIN enrollment prompt (see Figure \ref{fig:create:v3}) saying ``You won't need your PIN to open the app,'' many of the participants who did not set a PIN mentioned inconvenience as a reason for their decision. 

When further exploring the cause for this, the name ``PIN'' itself, is likely causing confusion. The Signal PIN is fundamentally a countermeasure against account takeover and to offer recovery functionality.  If for example, the Signal PIN were to be called the ``Account Recovery Password,'' or perhaps ``Restore/Recovery Password,'' that might better convey the usage pattern. Text could then inform the user of the ill consequences of a bad PIN choice. This end could be achieved with text like ``This password protects you from account takeover.''  Re-framing the PIN in this way could break the users' mental association with device-unlock PINs while also inspiring dread of consequences. While our study does not directly measure the effectiveness of such an intervention, the themes we uncovered naturally point in this direction.

%\figpromptsthird{}

\paragraph{Encouraging Password Managers}

The Signal PIN ultimately is used to derive a symmetric key in SVR and to retrieve a copy of the encrypted profile backup.  For this reason alone, it is worth encouraging users to generate and store their Signal PIN in a password manager (PM). Few users are willing to memorize long, random keys and a PM is much better at generation, storage, and recall of secrets. Importantly, the user interface of a PM is already designed to explain these concepts to a user. The longest and most diverse PINs observed in the data were selected by participants using a PM.

But to reach this goal, broader adoption of PMs is also needed: while half of the enthusiasts in our study are already using a PM to manage their Signal PIN, only 10 casual participants do (10\,\%).  For at least this group of users, this approach is preferable.  The Signal app could reinforce this idea in the UI and encourage users to adopt a PM if they have not yet~---~and if they have, to use it to manage their Signal PIN.

\paragraph{PIN Security}

An account with a strong PIN is less likely to be taken over by an attacker on the network.
%(an attacker with full control of a user's device can simply read the PIN when the user enters it).
Our data show large differences in how subgroups of participants select PINs.  Although we did not ask participants for their Signal PIN, we asked for its composition among classes of characters: digits, letters, and special characters.   Importantly Signal PIN security affects all users because account takeover can affect both the sender and receivers, especially in a group conversation. Even if a given user picks a strong PIN, if one of their messaging partners does not~---~that well-behaved user is at risk of mistakenly communicating sensitive data to an attacker who hijacked another account.

The current mechanisms of ensuring users select a strong PIN are minimal. Signal currently implements a very small blocklist of weak numeric PINs.  These include the following:
\begin{enumerate*}[label=(\alph*)]
    \item not empty;
    \item not sequential digits (e.g., 1234);
    \item not all the same digit (e.g., 0000)
\end{enumerate*}
Note that this leaves other popular choices like recent years and dates as acceptable Signal PINs, which are often chosen by users~\cite{bonneau-12-pin,markert-20-pin-blocklist}.  A targeted attack on an account where the victim's birthdate, anniversary, etc. are known would likely greatly assist the attacker.  The sequence check also only applies to numeric PINs~---~observe ``abcd'' and ``aaaa'' are both valid PINs.  In addition, this approach fails to block popular passwords like ``password.'' 

This situation could certainly be improved quite easily, for example implementing the blocklist as recommended by Markert et. al~\cite{markert-20-pin-blocklist} and Bonneau et. al~\cite{bonneau-12-pin} for PINs and following recent guidance from the literature and from government agencies for passwords. NIST Special Publication 800-63B, recommends checking user password choices against lists of the most popular passwords~\cite{nist-17-sp800-63b}. PIN checks could easily occur locally on the user's device; however full password checks would require additional features to protect the privacy of the user's password.

\paragraph{PIN Verification Reminders}

To our knowledge, this is the highest-profile roll-out to date of PIN verification reminders (both on Signal and other messengers using the Signal protocol, like WhatsApp).  While our study is based on user self-reported data, Figure~\ref{fig:PINMemorabilityAndVerification} shows that participants do not generally feel they have a problem recalling their Signal PIN. This could be due to password manager use or that participants are using PINs they know well and use in other contexts. More than half of users say they frequently/very frequently verify their PIN when prompted, which points to user acceptance of PIN reminders. Even though (45; 24\,\%) of respondents turned off PIN reminders, many of those used a password manager; the remainder appear to be comfortable and appreciate periodic PIN verification.

%% file: sections/07-conclusion.tex
%-------------------------------------------------------------------------------
\section{Conclusion}
%-------------------------------------------------------------------------------
\label{sec:conclusion}

We conducted an online study ($n=235$) of Signal users recruited from Reddit, Signal Community Forum, snowballing, and Prolific about their understanding and choice of Signal PINs. In total, 86\,\% of participants set a PIN, with 57\,\% able to technically describe what Signal PINs are used for (enthusiasts) and 43\,\% unable to accurately describe how Signal PINs are used (casuals). We also find that PIN composition followed similar lines: enthusiasts use significantly longer PINs with more complex compositions, and casual participants used more traditional, numeric PINs despite the fact that Signal allows PINs to be alphanumeric. This suggests that communication about the Signal PIN has been  effective for part of the Signal population only and that new strategies will be needed to reach the remainder.

As an example of in-app authentication~---~an authentication mechanism that occurs within a mobile app setting~---~our investigation shows that in the case of Signal, in-app usage of PINs can be confusing for users who have grown accustomed to screen lock and website login. These authentication metaphors are used often enough that users can be reasonably expected to handle them without much explanation.  Where some authentication machinery (a PIN, for example) is repurposed for symmetric-key derivation, only enthusiasts can be expected to read the blogs, documents, tweets, and online help text to gain a full understanding.

Thus, we conclude that communication needs to meet the understanding of the (possibly multiple) user communities.  Outside of a core constituency, even something as simple as the name matters.  Signal's choice of the term ``PIN'' can be seen as correct and well-understood by the developers and enthusiasts.  However, Signal may be well served in renaming their PIN, e.g., to ``Account Recovery Password,'' and other uses of in-app authentication will need to carefully choose names and messaging to match user expectations.  

Though our study does not measure the effect of this intervention, we believe there is strong evidence that suggests renaming Signal PIN to better reflect its usage could be helpful. First, a number of participants described it as an authentication mechanism or message privacy mechanism or simply indicated they do not know. A more precise name, like “Account Recovery,” would help users place the Signal PIN in context with other credentials they manage. Second, reusing the term ``PIN'' suggests to users that only digits are valid. Using the word ``Password'' or ``Passcode'' could elicit broader classes beyond digits and encourage more diverse composition.

%% file: sections/99-appendix.tex
\newpage
\section*{Appendix}

%\section{Demographics}
%\label{app:demo}
%\tableDemo{}

\section{Additional Pre-Screening Study}
\label{app:screen}
\begin{scriptsize}
The following question was asked in an additional pre-screening study on Prolific to be able to recruit more Signal users for our main study:

\begin{enumerate}[leftmargin=3em, label=\textbf{P{\arabic*}},noitemsep]
    \item Which instant messaging apps do you use? (Select all that apply) \newline
    $\square$~WhatsApp
    $\square$~Facebook Messenger
    $\square$~Signal
    $\square$~Telegram
    $\square$~iMessage
    $\square$~WeChat
    $\square$~QQ
    $\square$~Other, please specify: \rule{3.5cm}{.1pt}
\end{enumerate}
\end{scriptsize}

\section{Survey Instrument of the Main Study}
\label{app:survey}

%%%%%%%%%%%%%%%%%%%%%%%%%%%%%%%%%%%%%%%%%%%%%%%%%%%%%%%%%%%%%%%%%%%%%%%%%%%%%%%%

\definecolor{structure}{HTML}{03588C} % blue
\definecolor{note}{HTML}{BF2C47} % red
\newcommand{\questionspace}{\vspace{1em}}

\mbox{}
\setlength\parindent{0pt}
\begin{scriptsize}

\begin{center}
%    \vspace{-4em}
    \includegraphics[width=0.5\linewidth]{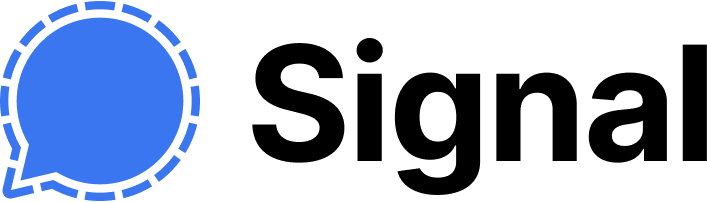}
%    \vspace{-1em}
\end{center}

\begin{enumerate}[leftmargin=3em, label=\textbf{Q\arabic*},noitemsep]

\item Signal Private Messenger is a cross-platform encrypted messaging service. Do you use Signal?\newline
$\circ$~Yes
$\circ$~No
\label{app:survey:q0}

\questionspace

\textit{{\color{note} [Participants who indicate No are screened out of the survey at this point, and only Signal users move forward]}}

\questionspace

\item I use Signal primarily on: \newline
$\circ$~Android
$\circ$~Apple iPhone \\
$\circ$~Other, please specify: \rule{3.5cm}{.1pt}
\label{app:survey:q1}

\questionspace

\item I also use Signal on: (Select all that apply)\newline
$\square$~Desktop
$\square$~Tablet
$\square$~None of these
\label{app:survey:q2}

\questionspace

\item PINs are a new feature provided by Signal.  In your own words, please explain how PINs are used by Signal. \newline
Answer: \rule{3.5cm}{.1pt}
\label{app:survey:q3}

\questionspace

\item Did you set a Signal PIN? \newline
$\circ$~Yes
$\circ$~No
\label{app:survey:q4}

%THE LABELS ARE Q1 Q4 etc .. not SQ ...fyi als you can referecnce
\questionspace

\textit{{\color{note} [Participants who indicate Yes to \ref{app:survey:q4}]}}\\
\vspace{-2.4em}

\end{enumerate}

\begin{enumerate}[leftmargin=3em, label=\textbf{Q6\alph*},noitemsep]
\item Why did you choose to set a PIN? \newline
Answer: \rule{3.5cm}{.1pt} \\
\label{app:survey:q5a} % all of these references need to be updated

\textit{{\color{note} [Participants who indicate No to \ref{app:survey:q4}]}}

\item Why did you choose not to set a PIN? \newline
Answer: \rule{3.5cm}{.1pt}
\label{app:survey:q5b} % all of these references need to be updated
\newline
\textit{{\color{note} [Participants who indicate No to \ref{app:survey:q4} skip ahead to \ref{app:survey:q25}]}}

\end{enumerate}

\begin{enumerate}[leftmargin=3em, label=\textbf{Q\arabic*},noitemsep]
\setcounter{enumi}{6}

\item Since setting your Signal PIN, are you still using it, or have you since disabled it? \newline
$\circ$~My Signal PIN is currently enabled
$\circ$~My Signal PIN is currently disabled

\label{app:survey:q6}

\questionspace

\textit{{\color{note} [Participants who indicated that their PIN is disabled in \ref{app:survey:q6}]}}
\item Why did you disable your Signal PIN? \newline
Answer: \rule{3.5cm}{.1pt}
\newline
\textit{{\color{note} [Participants who indicated that their PIN is disabled in \ref{app:survey:q6} skip ahead to \ref{app:survey:q25}]}}

\label{app:survey:q7}

\questionspace

\textit{{\color{note} [Participants who indicated that their PIN is enabled in \ref{app:survey:q6}]}}
\item How frequently do you have difficulty remembering your Signal PIN? \newline
$\circ$~Very frequently
$\circ$~Frequently
$\circ$~Occasionally
$\circ$~Rarely
$\circ$~Very rarely
$\circ$~Never

\label{app:survey:q8}

\questionspace
\textit{{\color{note} [Participants who indicated that their PIN is enabled in \ref{app:survey:q6}]}}
\item If you were to forget your Signal PIN, what would you do? \newline
Answer: \rule{3.5cm}{.1pt}

\label{app:survey:q9}

\questionspace

\textit{{\color{note} [A screenshot of the Verify PIN prompt (see Figure~\ref{fig:verify})]}} \\

\item Have you seen this dialog in Signal? \newline
$\circ$~Yes
$\circ$~No

\label{app:survey:q10}

\questionspace

\textit{{\color{note} [Participants who indicated that they have seen the dialog in \ref{app:survey:q10}]}}
\item When prompted, how frequently do you verify your Signal PIN? \newline
$\circ$~Very frequently
$\circ$~Frequently
$\circ$~Occasionally
$\circ$~Rarely
$\circ$~Very Rarely
$\circ$~Never

\label{app:survey:q11}

\questionspace
\textit{{\color{note} [Participants who indicated that they have seen the dialog in \ref{app:survey:q10}]}}
\item Have you disabled Signal PIN reminders? \newline
$\circ$~Yes
$\circ$~No

\label{app:survey:q12}

\newpage

\textit{{\color{note} [Participants who indicated that they have seen the dialog in \ref{app:survey:q10} and that they have disabled reminders in \ref{app:survey:q12}:]}}

\item Why did you disable Signal PIN reminders? \newline
Answer: \rule{3.5cm}{.1pt}

\label{app:survey:q14}

\questionspace

\item Many smartphone users also unlock their phone using a PIN or passcode. Is your Signal PIN the same one you use to unlock your smartphone? \newline
$\circ$~Yes
$\circ$~No
$\circ$~Unsure
$\circ$~I do not lock my smartphone with a PIN or passcode

\label{app:survey:q15}

\questionspace

\item Do you use your Signal PIN in other contexts besides unlocking your smartphone? (Select all that apply)\newline
$\square$~ATM/Credit/Payment Card
$\square$~Laptop/PC
$\square$~Online Accounts \\
$\square$~Electronic Door Lock
$\square$~Home Security System/Safe
$\square$~Garage Door Opener
$\square$~Car/Truck/SUV
$\square$~Bike/Gym lock
$\square$~Voicemail
$\square$~Gaming Console \\
$\square$~Smartwatch
$\square$~Other, please specify: \rule{3.5cm}{.1pt}

\label{app:survey:q16}

\questionspace

\item Do you use your Signal PIN in any other mobile applications? \newline
$\circ$~Yes, please specify: \rule{3.5cm}{.1pt}
$\circ$~No

\label{app:survey:q17}

\questionspace

\item Do you share your Signal PIN with friends or family? \newline
$\circ$~Yes
$\circ$~No

\label{app:survey:q18}

\questionspace

\item How long is your Signal PIN? \newline
Answer: \rule{3.5cm}{.1pt}

\label{app:survey:q19}

\questionspace

\item What was your primary strategy in selecting your Signal PIN? \newline
Answer: \rule{3.5cm}{.1pt}

\label{app:survey:q20}

\questionspace

\item Compared to other PINs you use, did you try to pick a Signal PIN that was: \newline
$\circ$~The most secure PIN you use
$\circ$~About the same security as other PINs you use
$\circ$~Less secure than other PINs you use

\label{app:survey:q21}

\questionspace

\item Why did you choose a PIN with this security level? \newline
Answer: \rule{3.5cm}{.1pt}
\label{app:survey:q22}

\questionspace

\item What is the shape of a red ball? \newline
$\circ$~Red
$\circ$~Round
$\circ$~Blue
$\circ$~Square

\label{app:survey:q23}
\questionspace

\textit{{\color{note} [For each category, this question uses sliders so the user can choose a value between 0 and 12, or check the category's box for ``Not applicable:'']}}

\item My Signal PIN contains: \newline
Digits: \rule{1.5cm}{.1pt} \\
Letters: \rule{1.5cm}{.1pt} \\
Special characters: \rule{1.5cm}{.1pt}

\label{app:survey:q24}

\questionspace

\item Do you use other messenger services like: (Select all that apply)\newline
$\square$~Facebook messenger
$\square$~Skype
$\square$~Telegram
$\square$~WeChat
$\square$~WhatsApp \\
$\square$~Other, please specify: \rule{3.5cm}{.1pt}

\label{app:survey:q25}

\textit{{\color{note} [For the services above, place them in order of how often you use them:]}}

\questionspace
\item Besides Signal, did you set a PIN in one or more other messengers? \newline
$\circ$~Yes
$\circ$~No

\label{app:survey:q26}

\questionspace

\textit{{\color{note} [Participants who indicate Yes to \ref{app:survey:q26}]}}
\vspace{-1.5em}
\end{enumerate}

\begin{enumerate}[leftmargin=3em, label=\textbf{Q27\alph*},noitemsep]

\item Why did you set a PIN in the other messenger(s)? \newline
Answer: \rule{3.5cm}{.1pt}
\label{app:survey:q27a} % all of these references need to be updated

\questionspace

\textit{{\color{note} [Participants who indicate No to \ref{app:survey:q26}]}}

\item Why didn't you set a PIN in the other messenger(s)? \newline
Answer: \rule{3.5cm}{.1pt}
\label{app:survey:q27b} % all of these references need to be updated
\newline
\textit{{\color{note} [Participants who indicate No to \ref{app:survey:q26} skip ahead to \ref{app:survey:d1}
]}}

\questionspace

\textit{{\color{note} [Participants who indicate Yes to \ref{app:survey:q26}]}}

\end{enumerate}

\begin{enumerate}[leftmargin=3em, label=\textbf{Q\arabic*}]
\setcounter{enumi}{27}

\vspace{-2.9em}
\item In which other messenger(s) did you set a PIN? (Select all that apply) \newline
$\square$~Facebook Messenger
$\square$~Skype
$\square$~Telegram
$\square$~WeChat
$\square$~WhatsApp \\
$\square$~Other, please specify: \rule{3.5cm}{.1pt}
\label{app:survey:q28}

\questionspace
\textit{{\color{note} [Participants who indicate Yes to \ref{app:survey:q26}]}}
\vspace{-1.1em}
\item Did you re-use the same PIN with any of these other messengers? \newline
$\circ$~Yes
$\circ$~No

\label{app:survey:q29}

\textit{{\color{note} [Participants who indicate Yes to \ref{app:survey:q29}]}}

\end{enumerate}

\vspace{-2.9em}

\begin{enumerate}[leftmargin=3em, label=\textbf{Q30\alph*},noitemsep]

\item Why did you re-use the same PIN in another messenger? \newline
Answer: \rule{3.5cm}{.1pt}
\label{app:survey:q30a} % all of these references need to be updated

\questionspace

\textit{{\color{note} [Participants who indicate No to \ref{app:survey:q29}]}}

\item Why didn't you re-use the same PIN in another messenger? \newline
Answer: \rule{3.5cm}{.1pt}
\label{app:survey:q30b} % all of these references need to be updated
\newline

\end{enumerate}

\newpage

\begin{enumerate}[leftmargin=3em, label=\textbf{D\arabic*},noitemsep]

\item What is your age range? \newline
$\circ$~18-24
$\circ$~25-34
$\circ$~35-44
$\circ$~45-54
$\circ$~55-64
$\circ$~65-74
$\circ$~75 or older
$\circ$~Prefer not to say

\label{app:survey:d1}

\questionspace

\item With what gender do you identify? \newline
$\circ$~Male
$\circ$~Female
$\circ$~Non-Binary
$\circ$~Other
$\circ$~Prefer not to say

\label{app:survey:d2}

\questionspace

\item What is the highest degree or level of school you have completed? \newline
$\circ$~Some high school
$\circ$~High school
$\circ$~Some college
$\circ$~Trade, technical, or vocational training
$\circ$~Associate's Degree
$\circ$~Bachelor's Degree
$\circ$~Master's Degree
$\circ$~Professional Degree
$\circ$~Doctorate
$\circ$~Prefer not to say

\label{app:survey:d3}

\questionspace

\item What is your country of residence? \newline
\textit{{\color{note} [Drop-down all countries]}}

\label{app:survey:d4}

\vspace{.85em}

\item Does your educational background or job field involve IT? \newline
$\circ$~Yes
$\circ$~No
$\circ$~Prefer not to say

\label{app:survey:d5}

\end{enumerate}
\end{scriptsize}

%\pagebreak

%\clearpage

\onecolumn

\section{Codebooks}
\label{app:codebook}
\small
We have 10 open-ended questions in our study for which two coders independently coded all answers we received. The two coders compared and combined codes until they agreed.  For each question, $n$ depicts the number of responses. As a single response might receive multiple codes, the number of codes does not sum to $n$.  All codes of participant responses are shown below.%As usual, multiple codes can be assigned to a single response.
%and Cohen's kappa $\kappa$ the level of agreement among the coders. Note, each quote can be assigned to multiple codes.

\renewcommand{\arraystretch}{1.1}
\begin{table}[htbp]
\scriptsize
%    \footnotesize
    \caption{
    \small \ref{app:survey:q3}: ``\emph{PINs are a new feature provided by Signal. In your own words, please describe how PINs are used by Signal.}'' ($n = 235$)}\label{tab:codebook-q3}%
    \begin{subtable}[h]{\textwidth}
    \caption{Based on the answer to \ref{app:survey:q3}, 132 participants were classified as \emph{enthusiasts}. }\label{tab:codebook-q3-enthusiasts}
    \begin{tabularx}{\linewidth}{>{\raggedleft\arraybackslash}p{6.7em} c c p{26em} p{26em}}
    \toprule
    \textbf{Code Name} & \textbf{No.} &    \textbf{\% } &
    \textbf{Description} & \textbf{Sample from the Study} \\
    \midrule
    Backup  & 65 & $49\,\%$ & Participant mentions secure backup of settings and contacts but not messages & 
    \emph{``The PIN enables storing a backup of the user’s settings on the signal servers in an encrypted form.''} (P13) \\
    
    Encryption & 45 & $34\,\%$ & Participant mentions encryption based on the PIN & 
    \emph{``deriving a key to encrypt data stored on signals servers''} (P82) \\
    
    Contacts & 31 & $24\,\%$ & Participant mentions the backup of contact data & 
    \emph{``To secure contacts data saved on signal server with your own pin''} (P7) \\
    
    Registration & 23 & $17\,\%$ &  Participant mentions the registration lock & 
    \emph{``to prevent reregistration of an account for the same mobile phone number for a given amount of time''} (P91) \\
    
    Settings & \hspace{.5em}8 & \hspace{.5em}$6\,\%$ & Participant mentions the backup of settings &
    \emph{``For encrypted backups - on cloud storage - for the user settings and profile. Not the messages themselves.''} (P127) \\
    
    Keying & \hspace{.5em}7 & \hspace{.5em}$5\,\%$ & Participant mentions the keying of the PIN & 
    \emph{``They say it's part of a keying mechanism providing a non-phone-number value that allows secure storage and retrieval of contacts and social graph info across devices.''} (P2) \\
    
    Phone number & \hspace{.5em}6 & \hspace{.5em}$5\,\%$ & Participant mentions the intention of Signal to move away from the phone number as an identifier & 
    \emph{``I think for backup purposes and to later fade out the phone number as identifier.''} (P106) \\
    
    Profile & \hspace{.5em}4 & \hspace{.5em}$3\,\%$ & Participant mentions the backup of profile information & 
    \emph{``PINs are used for recovery of settings and profile information after re-installation of Signal app.''} (P54) \\
    
    Groups & \hspace{.5em}3 & \hspace{.5em}$2\,\%$ & Participant mentions the backup of group memberships & 
    \emph{``They are used to secure private information such as group membership and store it on the Signal server'} (P35) \\
    
    Anti-Cloud & \hspace{.5em}2 & \hspace{.5em}$2\,\%$& Participant expresses negative sentiment about the data being stored by Signal & 
    \emph{``they are used to secure data in acloud service that is beeing forced on users''} (P208) \\
    
    SVR & \hspace{.5em}1 & \hspace{.5em}$1\,\%$& Participant mentions Secure Value Recovery (SVR) & 
    \emph{``Secure Value Recovery''} (P222) \\
    
    \bottomrule
    \end{tabularx}
    \end{subtable}
    
    \begin{subtable}[h]{\textwidth}
    \scriptsize
%    \footnotesize
    \vspace{1em}
    \caption{Based on the answer to \ref{app:survey:q3}, 103 participants were classified as \emph{casuals}. }\label{tab:codebook-q3-casuals}
    \begin{tabularx}{\linewidth}{>{\raggedleft\arraybackslash}p{6.7em} c c p{26em} p{26em}}
    \toprule
    \textbf{Code Name} & \textbf{No.} & \textbf{\%} & 
    \textbf{Description} & \textbf{Sample from the Study} \\
    \midrule
    Don't Know & 57 & $55\,\%$ & Participant does not mention any terms that may indicate an understanding & 
    \emph{``I don't understand their purpose very well. I thought that they might be using the PIN system to verify the identity of the person using signal (if for instance someone unauthorized gained access to the phone), but the way that pin entry is optionally offered every few weeks doesn't align with such a purpose. as such, I have no idea what they're trying to accomplish.''} (P178) \\
    
    Messages & 21 & $20\,\%$ & Participant mentions the backup of messages & 
    \emph{``Secure backup of messages''} (P23) \\
    
    Unlock & 21 & $20\,\%$ & Participant mentions that the PIN is used to protect access to the app & 
    \emph{``Protect application from opening from an unlocked phone''} (P37) \\
    
    Security & \hspace{.5em}2 & \hspace{.5em}$2\,\%$ & Participant mentions security & 
    \emph{``Security somehow...'} (P7) \\
    
    Inconvenient & \hspace{.5em}1 & \hspace{.5em}$1\,\%$ & Participant mentions inconvenience & 
    \emph{``I have not tried it considering that it’d pop up for additional verification through the pin.''} (P212) \\
    
    %Insecure & \hspace{.5em}1 & \hspace{.5em}$1\,\%$ & Participant mentions insecurity & 
    %\emph{``Weaken security properties to easen usability''} (P192) \\
    
    \bottomrule
    \end{tabularx}
    \end{subtable}
\end{table}

\begin{table}[htbp]
\scriptsize
%    \footnotesize
    \caption{\small  Codes assigned to the answers of the participants for (\ref{app:survey:q5a}) and (\ref{app:survey:q5b}) on adopting a PIN.}\label{tab:codebook-q5}%
    \begin{subtable}[h]{\textwidth}
    \caption{\ref{app:survey:q5a}: ``\emph{Why did you choose to set a PIN?}'' ($n = 202$)}\label{tab:codebook-q5a}
    \begin{tabularx}{\linewidth}{>{\raggedleft\arraybackslash}p{7em} c c c c p{21em} p{24em}}
    \toprule
    & \multicolumn{2}{c}{\textbf{Enthusiasts}} & \multicolumn{2}{c}{\textbf{Casuals}} \\ 
    \textbf{Code Name} & \textbf{No.} & \textbf{\%} & \textbf{No.} & \textbf{\%} &\textbf{Description} & \textbf{Sample from the Study} \\
    \midrule
    Security  & \hspace{.5em}7 & \hspace{.5em}5\,\% & 26 & 25\,\% & Participant mentions security & 
    \emph{``i wanted some extra security''} (P50) \\
    
    Required & 25 & 19\,\% & \hspace{.5em}6 & \hspace{.5em}6\,\% & Participant mentions that there was no other choice & \emph{``I did not see an option to not set one''} (P121) \\
    
    Prompted & \hspace{.5em}8 & \hspace{.5em}6\,\% & 13 & 13\,\% & Participant mentions that Signal showed a prompt that suggested it  & 
    \emph{``cause signal asked me to do so''} (P78) \\
    
    Don't Know & \hspace{.5em}4 & \hspace{.5em}3\,\% & 16 & 16\,\% &  Participant does not mention any of the terms that indicate an understanding & 
    \emph{``So that people that get a hold of my phone would have greater difficulty accessing my messages.''} (P159) \\
    
    Annoying & 12 & \hspace{.5em}9\,\% & \hspace{.5em}6 & \hspace{.5em}6\,\% & Participant mentions the feature was annoying &
    \emph{``Because it kept hassling you with a pop up screen''} (P154) \\
    
    Registration & 14 & 11\,\% & \hspace{.5em}2 & \hspace{.5em}2\,\% & Participant mentions the registration lock & 
    \emph{``I chose to set a PIN to both set registration lock and to backup my contacts.''} (P51) \\
    
    Features & \hspace{.5em}8 & \hspace{.5em}6\,\% & \hspace{.5em}3 & \hspace{.5em}3\,\% & Participant mentions features without further defining them &
    \emph{``To be able to use the features that depend on a PIN''} (P111) \\
    
    No harm & \hspace{.5em}8 & \hspace{.5em}6\,\% & \hspace{.5em}3 & \hspace{.5em}3\,\% & Participant describes there being no drawbacks & 
    \emph{``No disadvantage doing so''} (P20) \\
    
    Trust & \hspace{.5em}4 & \hspace{.5em}3\,\% & \hspace{.5em}2 & \hspace{.5em}2\,\% & Participant expresses trust in Signal & 
    \emph{``I trusted the app and just did it when prompted.''} (P155) \\
    
    Privacy & \hspace{.5em}2 & \hspace{.5em}2\,\% & \hspace{.5em}3 & \hspace{.5em}3\,\% & Participant mentions valuing privacy & 
    \emph{``Because privacy is important to me and it's an added layer of it''} (P162) \\
    
    Contacts & \hspace{.5em}3 & \hspace{.5em}3\,\% & \hspace{.5em}0 & \hspace{.5em}0\,\%  & Participant mentions the backup of contact data & 
    \emph{``I want to be able to access contact data saved on signal server if I somehow can't access my current phone''} (P7) \\
    
    Comfort & \hspace{.5em}2 & \hspace{.5em}2\,\% & \hspace{.5em}0 & \hspace{.5em}0\,\% & Participant mentions feeling comfortable & 
    \emph{``Because it I felt comfortable with the trade-off. Picked a long passphrase rather than a four digit PIN.''} (P127) \\
    
    Encryption & \hspace{.5em}1 & \hspace{.5em}1\,\% & \hspace{.5em}1 & \hspace{.5em}1\,\% & Participant mentions encryption based on the PIN & 
    \emph{``For me it's okay to encrypt and store data on Signal's servers as I have no high threat model.''} (P87) \\
    
    Lock & \hspace{.5em}1 & \hspace{.5em}1\,\% & \hspace{.5em}0 & \hspace{.5em}0\,\% & Participant mentions locking apart from registration lock & 
    \emph{``basically to lock and to avoid sim hijacking''} (P19) \\
    
    \bottomrule
    \end{tabularx}
    \end{subtable}
    
    \begin{subtable}[h]{\textwidth}
    \scriptsize
%    \footnotesize
    \vspace{1em}
    \caption{\ref{app:survey:q5b}: ``\emph{Why did you choose not to set a PIN?}'' ($n = 33$)}\label{tab:codebook-q5b}
    \begin{tabularx}{\linewidth}{>{\raggedleft\arraybackslash}p{7em} c c c c p{21em} p{24em}}
    \toprule
    & \multicolumn{2}{c}{\textbf{Enthusiasts}} & \multicolumn{2}{c}{\textbf{Casuals}} \\ 
    \textbf{Code Name} & \textbf{No.} & \textbf{\%} & \textbf{No.} & \textbf{\%} &\textbf{Description} & \textbf{Sample from the Study} \\
    \midrule
    Inconvenient & 4 & 25\,\% & 3 & 18\,\% & Participant mentions inconvenience & 
    \emph{``I want to access my apps as seamless and fast as possible.''} (P212) \\
    
    Anti-Cloud & 7 & 44\,\% & 0 & \hspace{.5em}0\,\% & Participant expresses negative sentiment about the data being stored by Signal & 
    \emph{``had no desire to have any contact data uploaded''} (P216) \\
    
    Key management & 3 & 19\,\% & 0 & \hspace{.5em}0\,\% & Participant described the use of the PIN in key derivation & 
    \emph{``I don't trust Signal's encryption strategy involving SGX. It's my belief that SGX is likely to be compromised by nation-state actors, and cannot be used securely. If any of my private information must be stored persistently in a cloud service, it is unacceptable to use anything other than an encryption key that I personally control.''} (P203) \\
    
    Lock & 0 & \hspace{.5em}0\,\% & 4 & 24\,\% & Participant falsely links the phone lock to the PIN & 
    \emph{``My phone is always locked. Additional authentication seems unnecessary''} (P227) \\
    
    No need & 0 & \hspace{.5em}0\,\% & 3 & 18\,\% & Participant mentions seeing no need & 
    \emph{``it's not necessary for me''} (P233) \\
    
    Memorability & 1 & \hspace{.5em}6\,\% & 1 & \hspace{.5em}6\,\% & Participant described memorability issues & 
    \emph{``I didn't want to be bothered with remembering another code.''} (P224) \\
    
    No awareness & 1 & \hspace{.5em}6\,\% & 0 & \hspace{.5em}0\,\% & Participant did not know Signal had a PIN & 
    \emph{``I didn't know it existed.''} (P232) \\
    
    Not prompted & 0 & \hspace{.5em}0\,\% & 1 & \hspace{.5em}6\,\% & Participant said they were not prompted to set a PIN & 
    \emph{``was not asked.''} (P218) \\
    
    Rarely use & 0 & \hspace{.5em}0\,\% & 1 & \hspace{.5em}6\,\% & Participant described using Signal only rarely & 
    \emph{``I dont use signal much, its not for sensitive messages so dont need the extra security''} (P223) \\
    
    Unsupported & 0 & \hspace{.5em}0\,\% & 1 & \hspace{.5em}6\,\% & Participant described using an unsupported client & 
    \emph{``Not possible because of using a unsupported native client for SailfishOS''} (P220) \\
    
    \bottomrule
    \end{tabularx}
    \end{subtable}
\end{table}

\begin{table}[htbp]
\scriptsize
%    \footnotesize
    \caption{\small \ref{app:survey:q7}: ``\emph{Why did you disable your Signal PIN?}'' ($n = 11$)}\label{tab:codebook-q7}%
    \begin{tabularx}{\linewidth}{>{\raggedleft\arraybackslash}p{7em} c c c c p{21em} p{24em}}
    \toprule
    & \multicolumn{2}{c}{\textbf{Enthusiasts}} & \multicolumn{2}{c}{\textbf{Casuals}} \\ 
    \textbf{Code Name} & \textbf{No.} & \textbf{\%} & \textbf{No.} & \textbf{\%} &\textbf{Description} & \textbf{Sample from the Study} \\
    \midrule
    Annoying & 3 & 60\,\% & 1 & 17\,\%  & Participant mentions being annoyed  & 
    \emph{``It was annoying.''} (P188) \\
    
    Anti-Cloud & 2 & 40\,\% & 1 & 17\,\%  & Participant expresses negative sentiment about the data being stored by Signal  & 
    \emph{``Don't want my data stored on their server''} (P193) \\
    
    Inconvenient  & 1 & 20\,\% & 1 & 17\,\%  & Participant mentions inconvenience & 
    \emph{``Verification overhead''} (P212) \\
    
    No backup & 1 & 20\,\% & 0 & \hspace{.5em}0\,\%  & Participant describes not needing a backup  &
    \emph{``It's annoying to re-enter the PIN and I don't need backup for signal since there's no important conversation''} (P231) \\
    
    No need & 1 & 20\,\% & 1 & 17\,\%  & Participant sees no necessity & 
    \emph{``I do not need it''} (P206) \\
    \bottomrule
    \end{tabularx}
\end{table}

\begin{table}[htbp]

\scriptsize
%    \footnotesize
\vspace{-.2in}
    \caption{\small \ref{app:survey:q9}: ``\emph{If you were to forget your Signal PIN, what would you do?}'' ($n = 191$) }\label{tab:codebook-q9}%
    \begin{tabularx}{\linewidth}{>{\raggedleft\arraybackslash}p{6em} c c c c p{23em} p{23em}}
    \toprule
    & \multicolumn{2}{c}{\textbf{Enthusiasts}} & \multicolumn{2}{c}{\textbf{Casuals}} \\ 
    \textbf{Code Name} & \textbf{No.} & \textbf{\%} & \textbf{No.} & \textbf{\%} &\textbf{Description} & \textbf{Sample from the Study} \\
    \midrule
    Don't know & 27 & 25\,\% & 33 & 40\,\%  & Participant does not know what to do  & 
    \emph{``Honestly don't know''} (P68) \\
    
    PW Manager  & 45 & 42\,\% & 12 & 15\,\%  & Participant has the PIN stored in a password manager & 
    \emph{``I've stored my Signal PIN in my PW manager''} (P74) \\
    
    Reset & \hspace{.5em}0 & \hspace{.5em}0\,\% & 12 & 15\,\%  & Participant describes resetting the account & 
    \emph{``Check the help page for how to reset''} (P158) \\
    
    Wait & \hspace{.5em}8 & \hspace{.5em}7\,\% & \hspace{.5em}4 & \hspace{.5em}5\,\%  & Participant is aware that the PIN expires and would wait & 
    \emph{``wait for pin expiration''} (P161) \\
    
    New PIN & \hspace{.5em}4 & \hspace{.5em}5\,\% & \hspace{.5em}7 & \hspace{.5em}6\,\%  & Participant would set a new PIN & 
    \emph{``as long as I have access to my Signal account I can set a new PIN at any time''} (P18) \\
    
    New account & \hspace{.5em}4 & \hspace{.5em}4\,\% & \hspace{.5em}5 & \hspace{.5em}6\,\%  & Participant would create a new account & 
    \emph{``I would make another account''} (P181) \\
    
    Reused & \hspace{.5em}2 & \hspace{.5em}2\,\% & \hspace{.5em}4 & \hspace{.5em}5\,\%  & Participant reuses the PIN and does not expect to forget it & 
    \emph{``It is a PIN I use for my bank cards, so I would not forget it.''} (P145) \\
    
    Unrecoverable & \hspace{.5em}2 & \hspace{.5em}2\,\% & \hspace{.5em}3 & \hspace{.5em}4\,\%  & Participant accepts that there is not way to recover & 
    \emph{``Signal said there is no way to recover it. All chats constants block list will be lost.''} (P79) \\
    
    Contact & \hspace{.5em}0 & \hspace{.5em}0\,\% & \hspace{.5em}4 & \hspace{.5em}5\,\%  & Participant would contact Signal directly &
    \emph{``Contact the signal team''} (P137) \\
    
    Guess & \hspace{.5em}0 & \hspace{.5em}0\,\% & \hspace{.5em}3 & \hspace{.5em}4\,\%  & Participant would try to guess the PIN & 
    \emph{``try a lot of PINs i use''} (P98) \\
    
    Reinstall & \hspace{.5em}2 & \hspace{.5em}2\,\% & \hspace{.5em}0 & \hspace{.5em}0\,\%  & Participant would reinstall Signal & 
    \emph{``delete the app and reinstall it''} (P106) \\
    
    Written & \hspace{.5em}1 & \hspace{.5em}1\,\% & \hspace{.5em}1 & \hspace{.5em}1\,\%  & Participant mentions that the PIN has been written down & 
    \emph{``I would check the PIN on my journal, I wrote it down with all the passwords and the login infos.''} (P143) \\
    \bottomrule
    \end{tabularx}
\end{table}

\begin{table}[htbp]
\scriptsize
%    \footnotesize
    \caption{\small \ref{app:survey:q14}: ``\emph{Why did you disable Signal PIN reminders?}'' ($n = 45$) }\label{tab:codebook-q14}%
    \begin{tabularx}{\linewidth}{>{\raggedleft\arraybackslash}p{6em} c c c c p{23em} p{23em}}
    \toprule
    & \multicolumn{2}{c}{\textbf{Enthusiasts}} & \multicolumn{2}{c}{\textbf{Casuals}} \\ 
    \textbf{Code Name} & \textbf{No.} & \textbf{\%} & \textbf{No.} & \textbf{\%} &\textbf{Description} & \textbf{Sample from the Study} \\
    \midrule
    PW Manager  & 22 & 67\,\% & \hspace{.5em}1 & \hspace{.5em}8\,\%  & Participant has the PIN stored in a password manager & 
    \emph{``Because I have a password safe and do not need to remember the pIn''} (P49) \\
    
    Annoyed & \hspace{.5em}6 & 18\,\% & \hspace{.5em}5 & 42\,\%  & Participant describes being annoyed  & 
    \emph{``Because it asked my pin to often''} (P70) \\
    
    No need & \hspace{.5em}5 & 15\,\% & \hspace{.5em}4 & 33\,\%  & Participant describes not needing them & 
    \emph{``I dont think I need them''} (P160) \\
    
    Memorized & \hspace{.5em}0 & \hspace{.5em}0\,\% & \hspace{.5em}1 & \hspace{.5em}8\,\%  & Participant does not expect to forget the PIN & 
    \emph{``Thought I’d be able to remember it''} (P157) \\
    
    Effective & \hspace{.5em}0 & \hspace{.5em}0\,\% & \hspace{.5em}1 & \hspace{.5em}9\,\%  & Participant mentions the effectiveness of the reminders  &
    \emph{``After a few reminders I was sure not to forget the PIN''} (P87) \\
    \bottomrule
    \end{tabularx}
\end{table}

\begin{table}[htbp]
\vspace{-.2in}
\scriptsize
%    \footnotesize
    \caption{\small \ref{app:survey:q20}: ``\emph{What was your primary strategy in selecting your Signal PIN?}'' ($n = 191$) }\label{tab:codebook-q20}%
    \begin{tabularx}{\linewidth}{>{\raggedleft\arraybackslash}p{6em} c c c c p{23em} p{23em}}
    \toprule
    & \multicolumn{2}{c}{\textbf{Enthusiasts}} & \multicolumn{2}{c}{\textbf{Casuals}} \\ 
    \textbf{Code Name} & \textbf{No.} & \textbf{\%} & \textbf{No.} & \textbf{\%} &\textbf{Description} & \textbf{Sample from the Study} \\
    \midrule
    Memorable  & 23 & 21\,\% & 30 & 36\,\%  & Participant mentions memorability & 
    \emph{``My ability to remember it.''} (P116) \\
    
    PW Manager & 28 & 26\,\% & \hspace{.5em}6 & \hspace{.5em}7\,\%  & Participant describes using a password manager  & 
    \emph{``My password safe generated it.''} (P100) \\
    
    Reuse & 16 & 15\,\% & 13 & 16\,\%  & Participant describes reusing a PIN & 
    \emph{``I used my PIN that I often use.''} (P176) \\
    
    Random & 15 & 14\,\% & \hspace{.5em}7 & \hspace{.5em}8\,\%  & Participant describes choosing a random PIN & 
    \emph{``random number generator''} (P63) \\
    
    Meaning & \hspace{.5em}6 & \hspace{.5em}6\,\% & \hspace{.5em}6 & \hspace{.5em}7\,\%  & Participant describes choosing a meaningful PIN &
    \emph{``Something meaningful to me''} (P77) \\
    
    Security & \hspace{.5em}3 & \hspace{.5em}3\,\% & \hspace{.5em}8 & 10\,\%  & Participant describes selecting a secure PIN & 
    \emph{``just something safe an long''} (P200) \\
    
    Pattern & \hspace{.5em}3 & \hspace{.5em}3\,\% & \hspace{.5em}4 & \hspace{.5em}5\,\%  & Participant describes choosing a PIN that depicts a pattern & 
    \emph{``Thinking of a pattern thats memorable to me''} (P142) \\
    
    None & \hspace{.5em}2 & \hspace{.5em}2\,\% & \hspace{.5em}3 & \hspace{.5em}4\,\%  & Participant describes not having a strategy & 
    \emph{``no strategy''} (P115) \\
    
    Word & \hspace{.5em}2 & \hspace{.5em}2\,\% & \hspace{.5em}3 & \hspace{.5em}4\,\%  & Participant describes converting a word to a PIN (textonyms) & 
    \emph{``Words to numbers''} (P115) \\
    
    Date & \hspace{.5em}2 & \hspace{.5em}2\,\% & \hspace{.5em}1 & \hspace{.5em}1\,\%  & Participant describes using a date & 
    \emph{``It‘s a date that is relevant but nobody knows''} (P154) \\
    
    System & \hspace{.5em}2 & \hspace{.5em}2\,\% & \hspace{.5em}1 & \hspace{.5em}1\,\%  & Participant describes having a certain system & 
    \emph{``My prefered format''} (P138) \\
    
    Typable & \hspace{.5em}0 & \hspace{.5em}0\,\% & \hspace{.5em}1 & \hspace{.5em}1\,\% & Participant mentions a PIN that is easy to enter & 
    \emph{``Strong alphanumeric password that is secure enough but fairly easy to type on the phone, even if I couldn't paste it from password manager for some reason.''} (P54) \\
    
    Simple & \hspace{.5em}0 & \hspace{.5em}0\,\% & \hspace{.5em}1 & \hspace{.5em}1\,\% & Participant mentions simplicity & 
    \emph{``Something simple''} (P154) \\
    
    Phone & \hspace{.5em}0 & \hspace{.5em}0\,\% & \hspace{.5em}1 & \hspace{.5em}1\,\% & Participant mentions a phone number & 
    \emph{``Old phone number i can remembee''} (P108) \\
    
    \bottomrule
    \end{tabularx}
\end{table}

\begin{table}[htbp]
\scriptsize
%    \footnotesize
    \caption{\small \ref{app:survey:q22}: ``\emph{Why did you choose a PIN with this security level?}'' ($n = 191$) }\label{tab:codebook-q22}%
    \begin{tabularx}{\linewidth}{>{\raggedleft\arraybackslash}p{6em} c c c c p{23em} p{23em}}
    \toprule
    & \multicolumn{2}{c}{\textbf{Enthusiasts}} & \multicolumn{2}{c}{\textbf{Casuals}} \\ 
    \textbf{Code Name} & \textbf{No.} & \textbf{\%} & \textbf{No.} & \textbf{\%} &\textbf{Description} & \textbf{Sample from the Study} \\
    \midrule
    Memorability  & 12 & 11\,\% & 18 & 22\,\%  & Participant mentions memorability & 
    \emph{``Because I wanted it to be easy to remember.''} (P5) \\
    
    Enough & 16 & 15\,\% & 15 & 18\,\%  & Participant describes the security level being sufficient  & 
    \emph{``I think that's enough''} (P179) \\
    
    Security & 25 & 23\,\% & 20 & 24\,\%  & Participant mentions security & 
    \emph{``I am fairly security conscious''} (P44) \\
    
    Consistent & 11 & 10\,\% & \hspace{.5em}6 & \hspace{.5em}7\,\%  & Participant describes this level being the standard & 
    \emph{``Because I always choose this security level.''} (P109) \\
    
    Trade-off & \hspace{.5em}9 & \hspace{.5em}8\,\% & \hspace{.5em}2 & \hspace{.5em}2\,\%  & Participant describes some form of trade-off & 
    \emph{``trade-off between remembering and security''} (P60) \\
    
    Reuse & \hspace{.5em}7 & \hspace{.5em}7\,\% & \hspace{.5em}2 & \hspace{.5em}2\,\%  & Participant describes reusing a PIN &
    \emph{``The same as the iPhone passcode.''} (P144) \\
    
    PW manager & \hspace{.5em}6 & \hspace{.5em}6\,\% & \hspace{.5em}2 & \hspace{.5em}2\,\%  & Participant describes using a password manager & 
    \emph{``why not, if i can use a pw manager''} (P76) \\
    
    Don't know & \hspace{.5em}1 & \hspace{.5em}1\,\% & \hspace{.5em}6 & \hspace{.5em}7\,\%  & Participant cannot remember the strategy & 
    \emph{``I don't remember''} (P84) \\
    
    None & \hspace{.5em}2 & \hspace{.5em}2\,\% & \hspace{.5em}4 & \hspace{.5em}5\,\%  & Participant describes not having a strategy & 
    \emph{``no strategy''} (P88) \\
    
    Convenience & \hspace{.5em}3 & \hspace{.5em}3\,\% & \hspace{.5em}1 & \hspace{.5em}1\,\%  & Participant mentions convenience & 
    \emph{``Convience over security''} (P113) \\  
    Privacy & \hspace{.5em}2 & \hspace{.5em}2\,\% & \hspace{.5em}1 & \hspace{.5em}1\,\%  & Participant mentions privacy & 
    \emph{``The chats and contacts in Signal have a relatively high level of privacy, so it should be properly protected. Yet the pin is not as good as for example my computers encryption password but as good as my android encryption phrase.''} (P94) 
    \\
    
    Low-threat & \hspace{.5em}0 & \hspace{.5em}0\,\% & \hspace{.5em}2 & \hspace{.5em}2\,\%  & Participant sees little need for data security & 
    \emph{``The info isn't super important''} (P168) 
    \\
    
    Indifference & \hspace{.5em}2 & \hspace{.5em}2\,\% & \hspace{.5em}0 & \hspace{.5em}0\,\%  & Participant says the PIN is unimportant & 
    \emph{``Dont think that the pin is too important''} (P120) 
    \\
    
    Rarely use & \hspace{.5em}2 & \hspace{.5em}2\,\% & \hspace{.5em}0 & \hspace{.5em}0\,\%  & Participant described using the PIN only rarely & 
    \emph{``Unlike my smartphone unlock pin for example, I don't have to enter my Signal PIN frequently (never really, unless I set up a new smartphone) and thus had no problem with selecting a long and complicated PIN''} (P20) 
    \\
    
    Minimum & \hspace{.5em}1 & \hspace{.5em}1\,\% & \hspace{.5em}0 & \hspace{.5em}0\,\%  & Participant mentions a Signal requirement & 
    \emph{``Initially 6 digits were required.''} (P132) 
    \\
    \bottomrule
    \end{tabularx}
\end{table}

\begin{table}[htbp]
\scriptsize
%    \footnotesize
    \caption{\small Codes assigned to the answers of the participants for (\ref{app:survey:q27a}) and  (\ref{app:survey:q27b}) on setting a PIN in other messengers.}\label{tab:codebook-q27}%
    \begin{subtable}[h]{\textwidth}
    \caption{\ref{app:survey:q27a}: ``\emph{Why did you set a PIN in other messenger(s)?}'' ($n = 49$) }\label{tab:codebook-q27a}
    \begin{tabularx}{\linewidth}{>{\raggedleft\arraybackslash}p{6em} c c c c p{23em} p{23em}}
    \toprule
    & \multicolumn{2}{c}{\textbf{Enthusiasts}} & \multicolumn{2}{c}{\textbf{Casuals}} \\ 
    \textbf{Code Name} & \textbf{No.} & \textbf{\%} & \textbf{No.} & \textbf{\%} &\textbf{Description} & \textbf{Sample from the Study} \\
    \midrule
    Security  & 24 & 67\,\% & \hspace{.5em}8 & 62\,\%  & Participant mentions security & 
    \emph{``For more security, 2FA''} (P95) \\
    
    Prompted & \hspace{.5em}4 & 11\,\% & \hspace{.5em}1 & \hspace{.5em}8\,\%  & Participant mentions being prompted by the application &
    \emph{``Prompted to do so, and I understand the reasons why it is a good idea.''} (P196) \\
    
    Required & \hspace{.5em}4 & 11\,\% & \hspace{.5em}1 & \hspace{.5em}8\,\%  & Participant mentions that there was no other choice & 
    \emph{`Forced to set''} (P93) \\
    
    Feature & \hspace{.5em}2 & \hspace{.5em}6\,\% & \hspace{.5em}1 & \hspace{.5em}8\,\%  & Participant mentions being given the option to  & 
    \emph{``Because I could''} (P117) \\
    
    Don't know & \hspace{.5em}1 & \hspace{.5em}3\,\% & \hspace{.5em}2 & 16\,\%  & Participant doesn't address the question & 
    \emph{``Telegram''} (P48) \\
    
    Reuse & \hspace{.5em}1 & \hspace{.5em}3\,\% & \hspace{.5em}0 & \hspace{.5em}0\,\%  & Participant mentions reusing a PIN when possible & 
    \emph{``Since I already has a pin memorized, why not use it in other messengers''} (P50) \\
    
    \bottomrule
    \end{tabularx}
    \end{subtable}
    
    \begin{subtable}[h]{\textwidth}
    \scriptsize
%    \footnotesize
    \vspace{1em}
    \caption{\ref{app:survey:q27b}: ``\emph{Why didn't you set a PIN in other messenger(s)?}'' ($n = 131$) }\label{tab:codebook-q27b}
    \begin{tabularx}{\linewidth}{>{\raggedleft\arraybackslash}p{6em} c c c c p{23em} p{23em}}
    \toprule
    & \multicolumn{2}{c}{\textbf{Enthusiasts}} & \multicolumn{2}{c}{\textbf{Casuals}} \\ 
    \textbf{Code Name} & \textbf{No.} & \textbf{\%} & \textbf{No.} & \textbf{\%} &\textbf{Description} & \textbf{Sample from the Study} \\
    \midrule
    No feature & 20 & 35\,\% & 24 & 33\,\%  & Participant mentions not being able to set a PIN  & 
    \emph{``They don't have that option''} (P35) \\
    
    No need & 18 & 32\,\% & 12 & 17\,\%  & Participant describe that there is no necessity & 
    \emph{``Not required''} (P156) \\
    
    Not asked & 10 & 17\,\% & 20 & 28\,\%  & Participant describes not being asked to & 
    \emph{``Was not asked to''} (P67) \\
    
    Use rarely & \hspace{.5em}3 & \hspace{.5em}5\,\% & \hspace{.5em}4 & \hspace{.5em}6\,\%  & Participant describes only using them rarely & 
    \emph{``I don't use them often, if at all.''} (P51) \\
    
    Screen lock & \hspace{.5em}3 & \hspace{.5em}5\,\% & \hspace{.5em}2 & \hspace{.5em}3\,\%  & Participant describes that the phone lock is sufficient & 
    \emph{``The phone in itself has a pin''} (P194) \\
    
    Annoyed & \hspace{.5em}2 & \hspace{.5em}3\,\% & \hspace{.5em}1 & \hspace{.5em}2\,\%  & Participant describes being annoyed & 
    \emph{``They are inconvenient, do not know how, and I do not use them for secure messaging. My Signal is already password protected so a pin seems redundant.''} (P55) \\
    
    Insecure & \hspace{.5em}0 & \hspace{.5em}0\,\% & \hspace{.5em}2 & \hspace{.5em}3\,\%  & Participant describes that they don't use them for secure communication & 
    \emph{``Not intended for secure communication.''} (P62) \\
    
    Comfort & \hspace{.5em}1 & \hspace{.5em}2\,\% & \hspace{.5em}0 & \hspace{.5em}0\,\%  & Participant mentions feeling comfortable & 
    \emph{``comfort''} (P102) \\
    
    \bottomrule
    \end{tabularx}
    \end{subtable}
\end{table}

\begin{table}[htbp]
\scriptsize
%    \footnotesize
    \caption{\small Codes assigned to the answers of the participants for (\ref{app:survey:q30a}) and  (\ref{app:survey:q30b}) on reusing the Signal PIN in another messenger. }\label{tab:codebook-q30}%
    \begin{subtable}[h]{\textwidth}
    \caption{\ref{app:survey:q30a}: ``\emph{Why did you re-use the same PIN in another messenger?}'' ($n = 10$) }\label{tab:codebook-q30a}
    \begin{tabularx}{\linewidth}{>{\raggedleft\arraybackslash}p{6.5em} c c c c p{23em} p{22.5em}}
    \toprule
    & \multicolumn{2}{c}{\textbf{Enthusiasts}} & \multicolumn{2}{c}{\textbf{Casuals}} \\ 
    \textbf{Code Name} & \textbf{No.} & \textbf{\%} & \textbf{No.} & \textbf{\%} &\textbf{Description} & \textbf{Sample from the Study} \\
    \midrule
    Memorability  & 5 & 63\,\% & 2 & 100\,\%  & Participant mentions memorability & 
    \emph{``I was too lazy to memorize a new one... not good I know''} (P50) \\
    
    Messenger PIN & 2 & 25\,\% & 0 & \hspace{1em}0\,\%  & Participant mentions using a PIN for messengers &
    \emph{``Because I have one pin for messengers.''} (P5) \\
    
    Convenience & 1 & 12\,\% & 0 & \hspace{1em}0\,\%  & Participant mentions convenience & 
    \emph{``Convivence, but it was probably a poor decision, as WhatsApp is more vulnerable to a secret warant.''} (P196) \\
    \bottomrule
    \end{tabularx}
    \end{subtable}
    
    \begin{subtable}[h]{\textwidth}
    \scriptsize
%    \footnotesize
    \vspace{1em}
    \caption{\ref{app:survey:q30b}: ``\emph{Why didn't you re-use the same PIN in another messenger?}'' ($n = 37$) }\label{tab:codebook-q30b}
    \begin{tabularx}{\linewidth}{>{\raggedleft\arraybackslash}p{6.5em} c c c c p{23em} p{22.5em}}
    \toprule
    & \multicolumn{2}{c}{\textbf{Enthusiasts}} & \multicolumn{2}{c}{\textbf{Casuals}} \\ 
    \textbf{Code Name} & \textbf{No.} & \textbf{\%} & \textbf{No.} & \textbf{\%} &\textbf{Description} & \textbf{Sample from the Study} \\
    \midrule
    Security & 17 & 65\,\% & 6 & 60\,\%  & Participant mentions security & 
    \emph{``Reusing PINs is a bad practice.''} (P54) \\
    
    PW Manager & \hspace{.5em}8 & 31\,\% & 2 & 20\,\%  & Participant describes using a password manager & 
    \emph{``Why would i? Thats what passwordmanagers are for.d''} (P23) \\
    
    Other options & \hspace{.5em}3 & 12\,\% & 0 & \hspace{.5em}0\,\%  & Participant describes having other options & 
    \emph{``Some of them gave me the option of using my thumbprint.''} (P3) \\
    
    Don't know & \hspace{.5em}1 & \hspace{.5em}4\,\% & 1 & 10\,\%  & Participant cannot explain the reason & 
    \emph{``I didn't really think about it, it just happened''} (P179) \\
    
    Required & \hspace{.5em}0 & \hspace{.5em}0\,\% & 1 & 10\,\%  & Participant mentions different requirements & 
    \emph{``different lengths''} (P381) \\
    \bottomrule
    \end{tabularx}
    \end{subtable}
\end{table}

\clearpage
\twocolumn
\section{Additional Figures}
\figpromptsall{}

%\section{Prompts Used by Signal}
%\label{app:prompts}